# Fragile-to-Fragile Liquid Transition at $T_g$ and Stable-Glass Phase Nucleation Rate Maximum at the Kauzmann Temperature $T_K$

**Robert F. Tournier**

Centre National de la Recherche Scientifique, Université Joseph Fourier, Consortium de Recherches pour l'Emergence de Technologies Avancées, B.P. 166, 38042 Grenoble-Cedex 09, France; E-Mail : robert.tournier@creta.cnrs.fr; Tel: +33-(0)608-716-878; Fax : +33-(0)476-881-280

**Abstract:** An undercooled liquid is unstable. The driving force of the glass transition at $T_g$ is a change of the undercooled-liquid Gibbs free energy. The classical Gibbs free energy change for a crystal formation is completed including an enthalpy saving. The crystal growth critical nucleus is used as a probe to observe the Laplace pressure change $\Delta p$ accompanying the enthalpy change $-V_m \times \Delta p$ at $T_g$ where $V_m$ is the molar volume. A stable glass-liquid transition model predicts the specific heat jump of fragile liquids at $T \leq T_g$, the Kauzmann temperature $T_K$ where the liquid entropy excess with regard to crystal goes to zero, the equilibrium enthalpy between $T_K$ and $T_g$, the maximum nucleation rate at $T_K$ of superclusters containing magic atom numbers, and the equilibrium latent heats at $T_g$ and $T_K$. Strong-to-fragile and strong-to-strong liquid transitions at $T_g$ are also described and all their thermodynamic parameters are determined from their specific heat jumps. The existence of fragile liquids quenched in the amorphous state, which do not undergo liquid-liquid transition during heating preceding their crystallization, is predicted. Long ageing times leading to the formation at $T_K$ of a stable glass composed of superclusters containing up to 147 atoms, touching and interpenetrating, are evaluated from nucleation rates. A fragile liquid-liquid transition occurs at $T_g$ without stable-glass formation while a strong glass is stable after transition.

## 1. Introduction

Vitrification is often viewed as a freezing-in process of undercooled melts instead of a phase change because there is, up to now, no intrinsic energy saving driving the formation of a new vitreous phase. A melt is seen as being stable with a well-defined viscosity and a unique temperature for the free-volume disappearance at the Vogel-Fulcher-Tammann temperature, which is deduced from the thermal variation of relaxation time or viscosity. Such a description leads to a natural freezing without thermodynamic transition because it does not include any modification of Gibbs free energy associated with molar volume thermal variation change. Nevertheless, the existence of a first-order transition near $T_g$ in triphenyl-phosphite is associated with a liquid instability [1-3]. Local minima in the potential energy landscape related to various local positions of all atoms are also considered to explain the equilibrium properties of amorphous substances [4].

Various models are considering the glass-liquid transition at $T_g$ as having a thermodynamic origin. Each broken chemical bond is viewed as an elementary configurational excitation called configuron. Using the Doremus' model of viscosity, the entropies and enthalpies of formation of configurons are obtained using a fitting process of experimental viscosities [5,6]. The glass-liquid transition has recently been treated within configuron percolation theory as a percolation-type phase transition with formation of dynamical fractal structures near the percolation threshold [7]. Specific heat jumps have been predicted. The formation of percolation structures made of high-density atom configurations in the glassy state was earlier suggested [8]. The nucleation of icosahedral clusters was also reported. Evteev et al studied, by molecular dynamics, atomic mechanisms of pure iron vitrification and showed that it is ensured by the formation of a percolation cluster from mutually penetrating and contacting icosahedrons with atoms at vertices and centers [9-11]. A fractal structure of icosahedrons incompatible with translation symmetry plays the role of binding carcass of the glassy state [10]. Wool developed the twinkling fractal theory following these initial ideas and found an explanation of relaxation phenomena near the glass transition. Clusters grow with fractal structures at the correlation length from molecular type units such as icosahedral superclusters, that are more and more frozen when the temperature decreases and lead to a disordered material [12,13]. Berthier et al showed that a growing length scale is accompanying the glass transition [14]. The twinkling dynamics of polystyrene was captured via atomic force microscopy within its glass transition region. Successive two-dimensional heights reveal that percolated clusters have lifetimes depending on their size and exist for longer time scales at lower temperatures. The computed fractal dimensions are shown to be in agreement with the theory of the fractal nature of percolating clusters [15]. All these models describe a liquid-liquid transition with liquid structures also containing superclusters below and above the glass transition and even beyond the melting temperature $T_m$ [16-18]. The same structure is progressively transformed and frozen by percolation instead of a new liquid phase growth from a critical nucleus. Kirpatrick and Thirumalai proposed a theory for the structural glass transition that is based on using frozen density fluctuations surrounded by surface energy to characterize it [19]. A random first-order phase transition is expected and hidden below the glass transition. The Kauzmann temperature would be the true glass transition at equilibrium [20]. Angell showed how such a theory for fragile liquids fits into a phenomenological scheme covering all glass formers [21]. A criterion analogous to the Lindemann criterion of melting was established for the glass-liquid transition [22]. It could be due to a vibrational instability of atoms located in the lattice sites. The new liquid state obtained just below $T_g$ has been still described by "means of reverse Monte Carlo (based on neutron scattering data) and molecular dynamics simulations showing that metallic glasses are composed by tiny icosahedral-like clusters, most of which are touching and/or interpenetrating yielding a microstructure of polyicosahedral clusters that follow a specific sequence of magic numbers" [23].

The glassy state relaxes more and more enthalpy when the temperature decreases below $T_g$. Spin glasses also relax a heat maximum when their remnant magnetization is saturated after applying a high magnetic field below the phase transition [24]. There is no more relaxed energy when the magnetization is equal to zero at thermodynamic equilibrium. The relaxed enthalpy below $T_g$ seems to be saturated after a relaxation time depending on the observation temperature. Its maximum cannot be larger than the available enthalpy which is frozen between $T_g$ and the Kauzmann temperature $T_K$ where the liquid entropy excess against that of crystal goes to zero [25]. There is a need to predict the glass enthalpy reduction accompanying the stable-glass formation in this temperature window. In our model, a change of the liquid Gibbs free energy occurs at the glass transition. The change associated with the crystallization of an undercooled liquid above $T_g$ contains the classical contribution $\theta \times \Delta H_m/V_m$ but also an enthalpy saving equal to $\varepsilon_{ls} \times \Delta H_m/V_m$ where $\theta$ is equal to $(T-T_m)/T_m$, $T_m$ being the melting temperature, $\Delta H_m$ the fusion heat per mole and $V_m$ the molar volume. This quantity corresponds in metallic liquids to the difference of conduction electron Fermi energies in crystals and their melt. At the vitreous transition temperature $T_g$ and at lower temperatures, there is a transformation of $\varepsilon_{ls}$ in $\varepsilon_{lgs}$ with a decrease equal to $\Delta \varepsilon_{lg} = (\varepsilon_{ls}-\varepsilon_{lgs})$. The enthalpy saving formulation $\Delta \varepsilon_{lg} \times \Delta H_m/V_m$ of the equilibrium glass phase below $T_g$ has already been proposed and applied to $Pd_{43}Cu_{27}Ni_{10}P_{20}$ [26]. But now, the coefficients $\varepsilon_{ls}$ and $\varepsilon_{lgs}$ in fragile liquids can be directly determined without using the specific heat jump at $T_g$. This model is based on the homogeneous nucleation of new n-atom superclusters characterized by a surface energy and an energy saving $\Delta \varepsilon_{lg} \times \Delta H_m \times n/N_A$ driving the formation of a stable-glass phase and being proportional to n where $\Delta \varepsilon_{lg}(\theta)$ depends on n and $\theta$, $N_A$ being the Avogadro number. The Gibbs free energy change associated with the formation of these new superclusters is equal to $\Delta\varepsilon_{lg} \times \Delta H_m/V_m$ and does not contain the classical contribution $\theta \times \Delta H_m/V_m$ associated with crystallization. The nucleation rate of these new superclusters has a maximum at $T_K$ instead of $T_g$ as shown in part 10. The existence of a liquid-liquid transition at $T_g$ is confirmed without nucleation of these entities. An enthalpy is always relaxed below $T_g$ in all liquids. The formation of the stable-glass phase in fragile glasses is accompanied by an exothermic latent heat which does not exist in strong glasses. The glass transition in strong liquids is a true liquid–liquid transformation without hidden latent heat as already shown in various models [6-8,10-15,19-22].

The classical model of crystal nucleation is completed by adding an unknown enthalpy saving in the Gibbs free energy change equal to $-\varepsilon_{ls} \times \Delta H_m = -V_m \times \Delta p$ for a supercluster formation, where $\Delta p$ is a complementary Laplace pressure acting on the growth critical nucleus [27]. A new equation for a nucleus formation has been established and the new homogeneous nucleation temperature corresponds to a minimum value of the surface energy for each value of $\varepsilon_{ls}$. The energy saving coefficient $\varepsilon_{ls}$ is a function of $\theta^2$, as already shown in liquid elements [28-30]. The derivative of the Gibbs free energy change $(-d\Delta G_{ls}/dT)_{Tm}$ for a critical nucleus formation at the melting temperature $T_m$ is equal to the bulk fusion entropy $\Delta S_m$. The coefficient $\varepsilon_{ls}$ has a maximum value $\varepsilon_{ls0}$ at $T = T_m$ which tends to zero at $T = T_{0m}$. The minimas of $\varepsilon_{ls0}$ and $\varepsilon_{ls}$ occur at a homogeneous nucleation temperature $\theta$ equal to $(\varepsilon_{ls} - 2)/3$. They depend on $\theta_{0m}$ (or $T_{0m}$) when the quenched liquid has escaped crystallization which is induced at higher temperatures by a tiny intrinsic nucleus reducing the effective energy barrier for crystal growth [29]. The change of $\varepsilon_{ls}$ in $\varepsilon_{lgs}$ at $T_g$ with $\varepsilon_{ls} > \varepsilon_{lgs}$, leads, far below the crystallization temperature, to the existence of two homogeneous nucleation temperatures $T_1$ and $T_2$, corresponding to two values $\varepsilon_{ls}$ and $\varepsilon_{lgs}$ above and below $T_g$ respectively, and to a change of $T_{0m}$ (or $\theta_{0m} = (T_{0m}-T_m)/T_m$)) in $T_{0g}$ (or $\theta_{0g} = (T_{0g}-T_m)/T_m$)) [31]. The homogeneous nucleation temperature $T_2$ (or $\theta_2$) equals $T_g$ (or $\theta_g$) because the liquid transformation has to lead to the minimum value of $\varepsilon_{lgs}$ [32,33]. When the transition takes places at a temperature lower or larger than $T_2$, the $\varepsilon_{lgs}$ is too large and leads to a relaxation towards its value at equilibrium. The coefficients $\varepsilon_{ls}$ and $\varepsilon_{lgs}$ tend to zero at the temperatures $T_{0m}$ and $T_{0g}$ with $T_{0m} > T_{0g}$ and are used as Vogel-Fulcher-Tammann temperatures $T_0$ characterizing the liquids above and below $T_g$. The transition at $T_g = T_2$ (or $\theta_g = \theta_2$) follows a scaling law that is a linear function of the energy saving coefficient $\varepsilon_{lgs0}$ of the new liquid phase extrapolated at $T_m$ [32]. The value of $T_g$ is used to determine $\varepsilon_{lgs}$ below $T_g$ and the temperature $T_{0g}$ (or $\theta_{0g}$), where $\varepsilon_{lgs}$ would be extrapolated to zero.

A temperature–time transformation (TTT) diagram describes the crystallization at temperatures much higher than $T_g$ with a nucleation time t of about 50-100 seconds at the nose temperature $T_n$ [34-36]. These crystallization temperatures are higher than the new homogeneous nucleation temperature $T_1$ calculated from the critical energy barrier because a tiny intrinsic nucleus reduces the effective energy barrier for crystal growth [28,37]. The isothermal nucleation total time t contains two added contributions $\pi^2/6 \times \tau^{ns}$ and $t_{sn}$, $\tau^{ns}$ being the time-lag for the transient nucleation and $t_{sn}$ the steady-state nucleation time [37]. The two contributions to t at the nose temperature $T_n$ being similar, $\tau^{ns}$ is of the order of 50 s at $T_n$ [29,31]. The time lag $\tau^{ns}$ is inversely proportional to K, as shown in (1), K being a coefficient in the exponential dependence of the supercluster nucleation rate J with the thermally-activated energy barrier $\Delta G_{eff}/k_BT$ in (2), and $\Gamma$ being the Zeldovich factor in (1) which is weakly dependent on the temperature, $N_A$ the Avogadro number and $k_B$ the Boltzmann constant [37]:

$$\tau^{ns} = \frac{\pi}{12\Gamma K} \frac{N_A}{V_m} \quad , \tag{1}$$

$$J = K \exp(-\frac{\Delta G_{eff}}{k_B T}). \qquad (2)$$

The measured isothermal relaxation time just below $T_g$ of a quenched melt can be viewed as being equal to $\tau^{ns}$ without including any contribution of $t_{sn}$ because only a liquid-liquid transition is considered. This is the time required for an equilibrium distribution of atoms to be established during the liquid-liquid transition preparing the steady-state nucleation of a vitreous phase [31,37]. The new liquid-phase formation is accomplished when the time-lag $\tau^{ns}$ is evolved. The coefficients K in (1-3), respectively called $K_{ls}$ at $T_n$ and $K_{lgs}$ at $T_g$, are nearly equal in this scheme in spite of their strong thermal dependence on the melt viscosity ratio $\ln(\eta/\eta_0) = B/(T-T_0)$ [38]:

$$\ln(K) = \ln(\frac{A\eta_0}{\eta}) = (\ln A) - \frac{B}{(T-T_0)}. \qquad (3)$$

Consequently, the A value in (3) is much stronger below $T_g$ than above $T_g$ [32]. There are two timescales above and below $T_g$. The time-lag above $T_g$, leading to a nucleus distribution ready for steady-state nucleation, is about $10^6$ times larger than the time-lag required below $T_g$. In Turnbull and Fisher's model, lnA is nearly equal to $\ln(N_A k_B T_g/V_m h) - \Delta f^*/k_B T_g$, where h is the Planck's constant and $\Delta f^*/k_B T^*_g$ a thermally-activated energy barrier for atom diffusion from the melt to the homogeneously-nucleated cluster [39]. This diffusion barrier $\Delta f^*/k_B T_g$ is reduced at $T_g$ during cooling. This expected change of activation enthalpy for diffusion was observed above and below $T_g$ in diffusivity measurements of impurities introduced in various glasses [40](Figure 12).

The stable-glass formation starts by homogeneous nucleation of condensed superclusters when the time lag $\tau^{ns}$ and their steady-state nucleation time $t_{sn}$ depending of their n-atom number are evolved. The energy barrier for growth being much too high, the stable phase is built by percolation and interpenetration of elementary superclusters containing magic atom numbers and accelerated by reduction of the interconnected supercluster surface energy. It is shown in appendix B that a TTT diagram of the stable vitreous phase exists below $T_g$ predicting the complete transformation of the liquid in stable phase with a minimum value of $t_{sn}$ for a maximum value of n occurring at the Kauzmann temperature.

The viscosity above $T_g$, in many examples of fragile glass-forming melts, perfectly obeys a scaling law, with a Vogel-Fulcher-Tammann (VFT) temperature equal to $0.77 \times T_g$ [41,42]. The homogeneous nucleation temperature $T_1$ in a fragile melt above $T_g$ calculated without any reduction of the critical energy barrier by a small homogeneously-condensed nucleus also follows a scaling law that is a linear function of $\varepsilon_{ls0}$. The comparison of theoretical and experimental scaling laws leads to the conclusion that the difference $\Delta\varepsilon_{lg0}$ between $\varepsilon_{lso}$ and $\varepsilon_{lgs0}$ in fragile liquids is equal to $-0.5 \times \theta_g$ [ [43]. The relaxed enthalpy maximum in the transformed liquid phase is equal to $\Delta\varepsilon_{lg0} \times \Delta H_m$ in strong and fragile liquids without including the exothermic formation heat of the stable-glass phase.

The model is able to determine the boundaries separating fragile from strong liquids [29,32]. The free-volume disappearance temperature $T_{0m}$ is less than $T_m/3$ in strong liquids and greater than $T_m/3$ in fragile liquids. The transition at $T_g$ is described by the formation at equilibrium of a new liquid phase characterized by an energy saving $\Delta\varepsilon_{lg} \times \Delta H_m$. The derivative $\Delta H_m \times d(\Delta\varepsilon_{lg})/dT$ is used to calculate the specific heat jump per mole at $T_g$ and the enthalpy saving varying from $T_g$ to the Kauzmann temperature $T_K$ where the liquid entropy excess compared to that of crystal goes to zero [25,26]. The equilibrium enthalpy change at $T_g$ is predicted. The presence or absence of equilibrium latent heat at $T_g$ is analyzed. The value of the Kauzmann temperature $T_K$ has already been determined in some glass-forming melts, observing that the specific heat jump between the undercooled liquid and the vitreous phase is nearly constant because the relaxed enthalpy has a nearly-linear decrease with the temperature increase [31,32]. The transition at $T_g$ leads to a much less fragile liquid with a temperature $T_{0g}$ lower than $T_{0m}$, and consequently in all cases to a stronger behavior below $T_g$. The following thermodynamic quantities are calculated: the coefficients $\varepsilon_{ls}$ and $\varepsilon_{lgs}$ above and below $T_g$, their difference $\Delta\varepsilon_{lg}(T)$, the specific heat jump $\Delta C_p(T_g)$, the temperatures $T_{0m}$ and $T_{0g}$ ($T_{0m} > T_{0g}$) at which the coefficients $\varepsilon_{ls}$ and $\varepsilon_{lgs}$ tend to zero in the undercooled and vitreous states respectively, the frozen enthalpy $\Delta H_g$ at $T_g$, the relaxed ultimate enthalpy $\Delta H_r$ and the frozen enthalpy below $T_g$ only knowing $T_g$, $\Delta H_m$ and the melting temperature $T_m$. The specific heat changes between $T_K$ and $T_g$ are predicted and used to determine the Kauzmann temperature of many fragile glass-forming melts.

About one third of fragile glass-forming melts do not follow the scaling law governing the viscosity above $T_m$. They are characterized by a larger energy saving $\Delta\varepsilon_{lg}$ leading to a larger specific heat jump. A reversible additional latent heat $L^-$ over that of liquids obeying scaling laws above $T_g$ is produced by heating and cooling through $T_g$. Values of the total latent heat $L^+$ obtained by heating, including the recovered relaxed enthalpy and $L^-$ when it exists, are proposed assuming that the new liquid properties continue to follow a universal scaling law below $T_g$ even when the energy coefficients for crystal nucleation are not separated by a universal value of $-0.5\times\theta_g$. The strong liquids have a specific heat jump $\Delta C_p(T_g)$ that is smaller than that of fragile liquids accompanying the decline of $T_{om}$. Their transition at $T_g$ occurs without latent heat during cooling and corresponds to $\Delta\varepsilon_{lg}(T_g) = 0$. Their specific heat jump at $T_g$ has to be known to determine $\theta_{0m}$, $\theta_{0g}$, $\Delta\varepsilon_{lg}$, $\varepsilon_{ls}$ and $\varepsilon_{lgs}$.

The effective critical energy barrier determined by the energy saving of stable vitreous clusters, being smaller than the homogeneous nucleation critical barrier, controls the nucleus growth and a possible phase change. This is not the first time that the state of the undercooled liquid has been viewed as being composed of long-lived structures created in the normal-liquid structure that is locally favored by a free energy decrease when the temperature decreases. These locally-favored structures may lead to a liquid-liquid phase transition [2,43]. The model developed in this paper and completed in Appendix A demonstrates that this transition occurs in all liquids at $T_g$ and that elementary superclusters are condensed at $T_K$ in fragile liquids after a very long ageing time, leading to a stable vitreous state composed of these numerous tiny entities instead of a single supercluster inducing the condensation of the whole vitreous phase because the critical effective nucleation barrier of the melt is too large to get over it. Thermodynamics shows that stable vitreous phases have to exist in all glasses. The recent discovery of ultra-stable glasses obtained by physical vapor deposition is a strong signal in favor of such analysis [44,45].

The TTT diagrams of three bulk metallic glass formers (BMG) in crystallized phases are reproduced in Appendix B calculating the effective thermally-activated energy barrier $\Delta G_{eff}/k_BT$ of the critical nucleus for crystal growth [29]. The critical nucleus is a 13-atom cluster having an effective energy barrier higher than that of bigger clusters. The condensation temperature of a single 13-atom cluster determines the crystallization temperatures of a whole sample along its TTT diagram. The energy saving coefficient of this type of cluster containing a small number of atoms embedded in glass-forming melts is quantified. The quantified value $\varepsilon_{13ls0}$ of a 13-atom nucleus deduced from the experimental TTT diagram is equal to 0.7, while those for 13-atom superclusters involved in stable-glass phase formation below $T_g$ are smaller.

The following plan is proposed:
2- Gibbs free energy change associated with growth nucleus formation,
3- Thermal dependence of the energy saving coefficient $\varepsilon_{nm}$ of an n-atom condensed cluster,
4- Critical supercluster homogeneous nucleation temperature and effective nucleation temperature,
5- The two homogeneous nucleation temperatures $T_1$ and $T_2$ and the equilibrium enthalpy change of glass-forming melt at the vitreous transition $T_g = T_2$,
6- Scaling laws,
7- The specific heat jump at fragile-to-fragile, strong-to-strong, and strong-to-fragile liquid transitions at $T_g$,
   *7.1- Fragile-to-fragile liquid transition,*
   *7.2- Strong-to-strong liquid transition,*
   *7.3-Strong-to-fragile liquid transition,*
8- Specific heat jumps from metallic and non-metallic glasses to undercooled liquids at the vitreous transition,
9- Enthalpy thermal cycles expected in some liquids and determination of the Kauzmann temperature,
10- Stable-glass supercluster nucleation rates between $T_K$ and $T_g$,
11 Fragile-to-fragile liquid transition at $T_g$ always occurring above $T_m/2$,
12- Hidden freezing at $T_g$ before relaxation of quenched liquids,
13- Transformation of the stable-glass Indomethacin in undercooled liquid at $T_g$,
14- Conclusions,
Acknowledgments,
Appendix A: TTT diagrams of several liquids in stable-glass phase between $T_K$ and $T_g$,
Appendix B: Superclusters of 13 atoms governing the first-crystallization time of metallic glass-forming melts.
References.

## 2. Gibbs free energy change associated with growth nucleus formation

The classical Gibbs free energy change for a nucleus formation in a melt is given by (4):

$$\Delta G_{1ls} = \theta \frac{\Delta H_m}{V_m} \frac{4\pi R^3}{3} + 4\pi \frac{\Delta H_m}{V_m} R^2 \sigma_{1ls}, \qquad (4)$$

where R is the nucleus radius and $\sigma_{1ls}$ is its surface energy. Turnbull has defined a surface energy coefficient $\alpha_{1ls}$ given by (5) which is equal to (6) [38,46]:

$$\sigma_{1ls}(\frac{V_m}{N_A})^{-1/3} = \alpha_{1ls}\frac{\Delta H_m}{V_m}, \qquad (5)$$

$$\alpha_{1ls} = \left[\frac{N_A k_B \ln(K_{ls})}{36\pi \Delta S_m}\right]^{1/3}. \qquad (6)$$

An energy saving per volume unit $-\varepsilon_{ls} \times \Delta H_m/V_m$ is introduced in (4) for a critical nucleus formation above $T_g$; the coefficient $\varepsilon_{ls}$ being replaced by $\varepsilon_{lgs}$ for a critical nucleus formation below $T_g$. The coefficients $\varepsilon_{ls}$ and $\varepsilon_{lgs}$ have to be calculated in order to determine their difference $\Delta\varepsilon_{lg}$ which determines the stable–glass formation enthalpy below $T_g$ when the quenched liquid escapes crystallization. The new Gibbs free energy change is given by (7), where $\sigma_{2ls}$ is the new surface energy [28]:

$$\Delta G_{2ls} = (\theta - \varepsilon_{ls})\frac{\Delta H_m}{V_m}\frac{4\pi R^3}{3} + 4\pi R^2 \sigma_{2ls}. \tag{7}$$

The new surface energy coefficient $\alpha_{2ls}$ is given by (8):

$$\sigma_{2ls}(\frac{V_m}{N_A})^{-1/3} = \alpha_{2ls}\frac{\Delta H_m}{V_m}. \tag{8}$$

The critical radius $R^*_{2ls}$ in (9) and the critical thermally-activated energy barrier $\Delta G^*_{2ls}/k_BT$ in (10) are calculated assuming $d\varepsilon_{ls}/dR = 0$:

$$R^*_{2ls} = \frac{-2\alpha_{2ls}}{\theta - \varepsilon_{ls}}(\frac{V_m}{N_A})^{1/3}, \tag{9}$$

$$\frac{\Delta G^*_{2ls}}{k_BT} = \frac{16\pi \Delta S_m \alpha_{2ls}^3}{3N_A k_B (\theta - \varepsilon_{ls})^2 (1+\theta)}. \tag{10}$$

They are no longer infinite at the melting temperature $T_m$ when $\varepsilon_{ls}$ is not equal to zero. The homogeneous nucleation temperature $T_2$ (or $\theta_2$) occurs when the nucleation rate J in (2) is equal to 1 and (11) is respected:

$$\frac{\Delta G^*_{2ls}}{k_BT} = \ln(K_{ls}). \tag{11}$$

The unknown surface energy coefficient $\alpha_{2ls}$ in (12) is deduced from (10) and (11):

$$\alpha_{2ls}^3 = \frac{3N_A k_B (\theta_2 - \varepsilon_{ls})^2 (1+\theta_2)\ln(K_{ls})}{16\pi \Delta S_m}. \tag{12}$$

The surface energy $\sigma_{2ls}$ in (8) has to be minimized to obtain the homogeneous nucleation temperature $T_1$ (or $\theta_1$) for a fixed value of $\varepsilon_{ls}$. The derivative $d\alpha_{2ls}/d\theta$ is equal to zero at the temperature $T_1$ (or $\theta_1$) given by (13) assuming that $\ln(K)$ does not depend on the temperature:

$$\frac{d\alpha_{2ls}^3}{d\theta} = 3\alpha_{2ls}^2 \frac{d\alpha_{2ls}}{d\theta} \approx (\theta - \varepsilon_{ls})(3\theta + 2 - \varepsilon_{ls}),$$

$$\theta_1 = \frac{T_1 - T_m}{T_m} = \frac{\varepsilon_{ls} - 2}{3} \quad . \tag{13}$$

The enthalpy saving between solid and liquid states is determined by the knowledge of $\theta_1$.

The thermal variation of $\ln(K_{ls})$, being a function of viscosity, does not modify the value of $T_1$ (or $\theta_1$) as shown below [29]. The critical energy barrier in (10) is the product of a function $g(\theta)$ and $\ln(K_{ls})$. The maximum nucleation rate J still occurs at the temperature $T_1$ with the derivative $d(\ln J)/d\theta$ being equal to zero because $g(\theta)$ is equal to 1 and $dg(\theta)/d\theta = 0$:

$$\ln J = \ln(K_{ls}) - \frac{\Delta G^*_{2ls}}{k_B T} = \ln(K_{ls}) - g(\theta)\ln(K_{ls}) \quad ,$$

$$\frac{d\ln J}{d\theta} = \frac{d\ln(K_{ls})}{d\theta} - \frac{dg(\theta)}{d\theta}\ln(K_{ls}) - \frac{d\ln(K_{ls})}{d\theta}g(\theta) = 0.$$

The surface energy coefficient $\alpha_{2ls}$ is now given by (14) replacing $\theta$ by (13) in (12) for each value of $\varepsilon_{ls}$:

$$\alpha_{2ls} = (1+\varepsilon_{ls})\left[\frac{N_A k_B \ln(K_{ls})}{36\pi \Delta S_m}\right]^{1/3} = (1+\varepsilon_{ls})\alpha_{1ls}. \tag{14}$$

The classical nucleation equation (4) has been transformed into (15) introducing the energy saving coefficient $\varepsilon_{ls}$:

$$\Delta G_{2ls}(R,\theta) = \frac{\Delta H_m}{V_m}(\theta - \varepsilon_{ls})4\pi\frac{R^3}{3} + 4\pi R^2 \frac{\Delta H_m}{V_m}(1+\varepsilon_{ls})(\frac{12k_B V_m \ln K_{ls}}{432\pi \times \Delta S_m})^{1/3}. \tag{15}$$

The Laplace pressure p can be calculated from the surface energy $\sigma_{2ls}$ with the equation (13) [32,33]:

$$p = \frac{2\sigma_{2ls}}{R} = \frac{\Delta H_m}{V_m}\left[\theta - \varepsilon_{ls}(\theta)\right]. \tag{16}$$

The complement $\Delta p$ of Laplace pressure introduced by the energy saving $\varepsilon_{ls}$ is equal to $\varepsilon_{ls}(\theta) \times \Delta H_m/V_m$. All equations (7-16) for a critical nucleus formation can be applied below $T_g$ after replacing $\varepsilon_{ls}$ by $\varepsilon_{lgs}$ and $T_1$ (or $\theta_1$) by $T_2$ (or $\theta_2$). The equations governing the stable-glass supercluster formation below $T_g$ are developed in section 10. The Gibbs free energy change $\Delta G_{2ls}$ in (15) directly depends on the cluster atom number n and the energy saving coefficient $\varepsilon_{nm}$ of the cluster instead of depending on its molar volume $V_m$ and its radius R, as shown in (17):

$$\Delta G_{nm}(n,\theta,\varepsilon_{nm}) = \Delta H_m \frac{n}{N_A}(\theta - \varepsilon_{nm}) + \frac{(4\pi)^{1/3}}{N_A}\Delta H_m \alpha_{2ls}(3n)^{2/3}. \tag{17}$$

This equation can be applied below and above $T_g$ using different values of $\varepsilon_{nm}$ corresponding to various Laplace pressures.

The formation of superclusters having a weaker energy barrier precedes the formation of crystallized nuclei in an undercooled melt [9,16,47,48]. A supercluster containing n atoms could be easily transformed into a crystal of n atoms having the same energy saving coefficient $\varepsilon_{nm}$ and the same surface without changing $\Delta G_{nm}$. The transformation of superclusters into crystals could occur when their molar volume and consequently their coefficient $\varepsilon_{nm}$ becomes equal to those of crystals. This condition could be not sufficient because crystals of n atoms could be facetted, less-spherical and their surface not minimized. The supercluster energy saving $\varepsilon_{nm} \times \Delta H_m$ is quantified, depending on the radius and atom number n, and can be smaller or larger than the critical energy saving $\varepsilon_{ls} \times \Delta H_m$ in glass-forming melts above $T_g$ and larger than the critical energy saving $\Delta\varepsilon_{lg} \times \Delta H_m$ of the stable-glass phase appearing below $T_g$. The critical growth barrier $\Delta G^*_{nm}$ of an n-atom supercluster also depends on $\varepsilon_{nm}$. It can be larger than the critical growth barrier $\Delta G^*_{2ls}$. The effective critical energy barrier of the smallest homogeneously-condensed cluster can control the heterogeneous growth around it. It is the case for the crystallization of glass-forming melts at temperatures higher than the homogeneous nucleation temperature defined by (13).

## 3. Thermal dependence of the energy saving coefficient $\varepsilon_{nm}$ of an n-atom condensed cluster

All superclusters which are formed in an undercooled melt preparing crystal formation grow first, are submitted to a complementary Laplace pressure and have a surface energy because the Gibbs free energy change contains an enthalpy saving [9,27,48]. The energy saving coefficient $\varepsilon_{nm}$ of an n-atom supercluster giving rise to crystals has to be a function of $\theta^2$ to survive above $T_m$, assuming that $\varepsilon_{nm}$ is maximum at $T_m$, $d\varepsilon_{nm}/dT$ being equal to zero and fixing the supercluster fusion entropy as being equal to the fusion entropy $\Delta S_m$ of the bulk solid [29]:

$$\frac{3}{4\pi R^3}\left[\frac{d(\Delta G_{nm})}{dT}\right]_{T=Tm} = \frac{-\Delta S_m}{V_m}.$$

In these conditions, the cluster fusion occurs above $T_m$ and is governed by liquid droplet homogeneous nucleation above $T_m$ rather than by surface melting. This $\theta^2$ thermal variation explains the undercooling rate of liquid elements [28,48].

The law (18) is expected to work also in glass-forming melts as observed in liquid elements, $\varepsilon_{nm}$ being the quantified energy saving coefficient of an n-atom supercluster:

$$\varepsilon_{nm} = \varepsilon_{nm0}\frac{\Delta H_m}{V_m} = \varepsilon_{nm0}(1-\frac{\theta^2}{\theta_{0m}^2})\frac{\Delta H_m}{V_m}. \tag{18}$$

A n-atom supercluster is submitted to a Laplace pressure which is strongly dependent of its radius. The quantified value $\varepsilon_{nm}$ decreases with $\theta^2$ and becomes equal to zero when the temperature equals $T_{0m}$ (or $\theta_{0m}$). An atom which does not belong to a supercluster cannot be involved in the nucleation of a new phase.

The critical parameters for supercluster growth are determined by an energy saving coefficient called $\varepsilon_{ls}$ (or $\varepsilon_{lgs}$) in (19):

$$\varepsilon_v = \varepsilon_{ls}\frac{\Delta H_m}{V_m} = \varepsilon_{ls0}(1-\frac{\theta^2}{\theta_{0m}^2})\frac{\Delta H_m}{V_m}. \tag{19}$$

The enthalpy saving per mole $\varepsilon_{ls}\times\Delta H_m$ tends to zero when $\theta$ tends to $\theta_{0m}$. The liquid free volume would be equal to zero at $T = T_{0m}$ in the absence of transformation at $T_g$.

A critical supercluster contains a critical number $n_c$ of atoms given by (20):

$$n_c = \frac{32\pi\alpha_{2ls}^3}{3(\theta-\varepsilon_{ls})^3}. \tag{20}$$

## 4. Critical supercluster homogeneous nucleation temperature and effective nucleation temperature

The thermally-activated critical energy barrier is now given by (21), where $\varepsilon_{ls}$ is given by (19):

$$\frac{\Delta G_{2ls}^*}{k_B T} = \frac{12(1+\varepsilon_{ls})^3 \ln(K_{ls})}{81(\theta-\varepsilon_{ls})^2(1+\theta)}. \tag{21}$$

The coefficient of $\ln(K_{ls})$ in (21), called $g(\theta)$, becomes equal to 1 at the homogeneous nucleation temperature given in (13) and the equation (11) is respected. Homogeneously-condensed clusters of n-atoms act as growth nuclei at a temperature generally higher than the homogeneous nucleation temperatures $T_1$ and $T_2$ of liquids. The cluster thermally-activated critical energy barrier $\Delta G^*_{nm}/k_B T$ and its effective thermally-activated critical energy barrier are given by (22):

$$\frac{\Delta G_{neff}}{k_B T} = \frac{\Delta G_{nm}^*}{k_B T} - \frac{\Delta G_{nm}}{k_B T} = \frac{12(1+\varepsilon_{nm})^3 \ln(K)}{81(\theta-\varepsilon_{nm})^2(1+\theta)} - \frac{\Delta G_{nm}}{k_B T}, \tag{22}$$

where $\Delta G_{nm}$ is given by (17) and $\varepsilon_{nm}$ by (18). These nuclei are ready to grow when the transient nucleation time $\tau^{ns}$ and the steady-state nucleation time $t_{sn}$ are evolved and the relation (23) respected:

$$\ln(J_n v t_{sn}) = \ln(K_{ls} v t_{sn}) - \frac{\Delta G_{neff}}{k_B T} = 0. \tag{23}$$

Three situations are met when the two critical energy barriers in (21) and (22) are compared. The growth around nuclei can be very rapid if $\Delta G^*_{2ls}/k_B T$ given in (21) is smaller than $\Delta G^*_{nm}/k_B T$ given in (22). The steady-state nucleation time $t_{sn}$ (T) is calculated with (23) knowing the sample volume v. This is the case for glass-forming melt crystallization above $T_g$. When $\Delta G^*_{2ls}/k_B T$ is larger than $\Delta G^*_{nm}/k_B T$, the new effective critical energy barrier is equal to ($\Delta G^*_{2ls}/k_B T-\Delta G_{nm}/k_B T$) and has to replace $\Delta G_{neff}/k_B T$ in (23). The effective heterogeneous nucleation temperature is strongly dependent on the volume v. This phenomenon is very important in liquid elements where the effective nucleation temperature is observed around θ = −0.2 in sample volumes of few mm$^3$ instead of θ varying from −0.58 to −0.3 with much smaller samples [48,49]. The growth around clusters can be slow and shell-by-shell as a function of time when $\Delta G^*_{2ls}/k_BT$ is much larger than $\Delta G^*_{nm}/k_BT$. It is the case for the glass phase nucleation critical barrier $\Delta G^*_{2lg}/k_BT$ which is much larger than the critical barrier associated with an elementary n-atom cluster $\Delta G^*_{nm}/k_BT$.

**5. The two homogeneous nucleation temperatures T$_1$ and T$_2$ and the equilibrium enthalpy change of glass-forming melt at the vitreous transition T$_g$ = T$_2$**

The homogeneous nucleation temperature T$_1$ (or θ$_1$) obeys the equations (13) and (19) and is a solution of the equation (24):

$$\frac{\theta_1^2 \times \varepsilon_{ls0}}{\theta_{0m}^2} + 3\theta_1 + 2 - \varepsilon_{ls0} = 0, \tag{24}$$

where θ$_{0m}$ (or T$_{0m}$) is the temperature corresponding to ε$_{ls}$ = 0.
At the vitreous transition T$_g$, there are changes of ε$_{lso}$ in ε$_{lgs0}$, θ$_1$ in θ$_2$ (or T$_1$ in T$_2$) and θ$_{0m}$ in θ$_{0g}$ (or T$_{0m}$ in T$_{0g}$). A new equation (25) is used to calculate θ$_2$ (and T$_2$):

$$\frac{\theta_2^2 \times \varepsilon_{lgs0}}{\theta_{0g}^2} + 3\theta_2 + 2 - \varepsilon_{lgs0} = 0. \tag{25}$$

There are two values of θ$_1$ and θ$_2$ respecting (24) or (25) for each value of θ$_{0m}$ or θ$_{0g}$; the largest value given in (26) is the chosen solution for θ$_1$ and θ$_2$:

$$\theta_1 = \frac{-3 \pm \left[9 - 4(2 - \varepsilon_{ls0})\varepsilon_{ls0}/\theta_{0m}^2\right]^{1/2}}{2\varepsilon_{ls0}} \theta_{0m}^2. \tag{26}$$

The fragility of melts defined by Angell is larger when θ$_{0m}$ increases [50]. In strong liquids, the temperature θ$_1$ can be calculated using (26) when ε$_{ls0}$ ≤ 2 and θ$_{0m}$ ≤ −2/3 are known. There is a double solution for a fragile liquid and a minimum value of ε$_{ls0}$ ≥ 1 for each value of θ$_{0m}$ ≥ −2/3 (or T$_{0m}$ > T$_m$/3) [ [51]]. The boundary separating strong from fragile liquids is θ$_{0m}$ = −2/3 (or T$_m$/3). The values of θ$_1$ and θ$_{0m}$ in fragile liquids are given as a function of ε$_{ls0}$ in (27) and (28) with "a" being unknown:

$$\varepsilon_{ls0} = \varepsilon_{ls}(\theta = 0) = 1.5 \times \theta_1 + 2 = a \times \theta_g + 2, \tag{27}$$

$$\theta_{0m}^2 = \frac{8}{9}\varepsilon_{ls0} - \frac{4}{9}\varepsilon_{ls0}^2. \tag{28}$$

The thermodynamic transition at T$_g$ is induced by a liquid-liquid transformation at the temperature T$_2$ (or θ$_2$) minimizing the energy saving ε$_{lgs}$ of the new liquid state for a given value of θ$_{0g}$. Any other transition temperature leads in fragile liquids to a too large value of ε$_{lgs0}$ and to an energy relaxation. The thermodynamic transition temperature T$_g$ has to be also equal to T$_2$. The thermodynamic parameters characterizing the new liquid state at equilibrium are $\varepsilon_{lgs0} = (1.5 \times \theta_g + 2)$ and $\theta_{0g}$, given in (27) and (28) after replacing $\varepsilon_{ls0}$ by $\varepsilon_{lgs0}$, $\theta_1$ by $\theta_g$ and $\theta_{0m}$ by using $\theta_{0g}$. The transformation of any liquid in strong liquid at $T_g$ will occur when the difference $\Delta\varepsilon_{lg} = (\varepsilon_{ls} - \varepsilon_{lgs})$ becomes equal to zero because there is no minimum of $\varepsilon_{lgs0}$ as a function of $\theta_2$ in strong liquids.

A crystallization at $T_g = T_2$ is not possible because the homogeneous nucleation rate of a critical supercluster is only equal to 1 m$^{-3}$.s$^{-1}$. In addition, it cannot start from a 13-atom supercluster because its own nucleation time is much too large. The energy saving coefficient, driving the stable glass phase formation in the new liquid state as a function of temperature, has to vary at thermodynamic equilibrium as $\Delta\varepsilon_{lg} = (\varepsilon_{ls} - \varepsilon_{lgs})$ as shown below. The enthalpy decrease per mole from quenched undercooled liquids, aged at temperatures less than or equal to $T_g$ to lead to stable vitreous states, is equal in (29) to the difference between the complementary energy savings for a crystal formation below $T_g$ [26] :

$$\Delta H = -\Delta H_m \times \Delta\varepsilon_{lg} = -\Delta H_m \left[ \varepsilon_{ls0} - \varepsilon_{lgs0} + \theta^2 \left( \frac{\varepsilon_{lgs0}}{\theta_{0g}^2} - \frac{\varepsilon_{ls0}}{\theta_{0m}^2} \right) \right] . \qquad (29)$$

This difference $\Delta\varepsilon_{lg}$ occurring at $T_g$ and below $T_g$ is only due to the change of the equilibrium enthalpy between $T_g$ and the Kauzmann temperature $T_K$. It contains a temperature-dependent positive contribution and a constant negative contribution which is viewed as the ultimate relaxed enthalpy at $T_K$ for quenched liquids respecting the scaling laws above $T_g$. The stable-glass-to-fragile liquid transformation of indomethacin at equilibrium is accompanied by an endothermic latent heat at $T_g$, as shown in Figure 1. A strong-glass-to-strong liquid or a strong-glass-to-fragile liquid transformation occurs without latent heat when $\Delta\varepsilon_{lg} = 0$ instead of being due to kinetic effects [32].

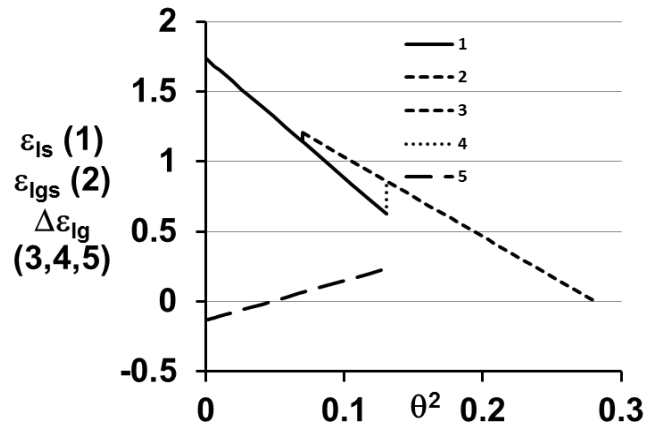

**Figure 1. Enthalpy saving coefficients versus $\theta^2 = [(T-T_m)/T_m]^2$.** The enthalpy saving coefficients $\varepsilon_{ls}$ and $\varepsilon_{lgs}$ for critical nucleus formation, proportional to the complementary Laplace pressure $\Delta p$, for an indomethacin nucleus containing a critical atom number $n_c(T)$ respectively above and below the transition $\theta_g = -0.264$, are plotted versus $\theta^2$. The differences $(\varepsilon_{ls}-\varepsilon_{lgs}) \times \Delta H_m = \Delta\varepsilon_{lg} \times \Delta H_m$ at $T_g = 318$ K and at the Kauzmann temperature $T_K$ determines the equilibrium latent heats involved in the stable-glass-to–fragile-liquid transition at $T_g$ and $T_K$. The coefficient $\Delta\varepsilon_{lg}$ given in (29) is plotted versus $\theta^2$ between $T_m$ and $T_K$.

An example is given in Figure 2 to illustrate the change of energy saving at $T_2 = T_g$ in fragile-to-fragile liquid transformations. The coefficients $\varepsilon_{ls0}$ and $\varepsilon_{lgs0}$ of the metallic glass-forming melt $Pd_{43}Ni_{10}Cu_{27}P_{20}$ are calculated using $T_{0m} = 430$ K and $T_{0g} = 365$ K as a function of the homogeneous nucleation temperature $T_2$. The energy saving coefficient $\varepsilon_{ls0}$ for a crystal formation is equal to 1.718 for $\theta_1 = -0.188$. The coefficient $\varepsilon_{lgs0}$ has a minimum equal to 1.577 at $\theta_2 = -0.282$. All values of $\varepsilon_{lgs0}$ below and above $\theta_2 = -0.282$ are larger. A liquid quench down to $T_a$ is followed by enthalpy relaxation from $\varepsilon_{ls0} = 1.718$ down to $\varepsilon_{lgs0} = 1.577$. The thermodynamic glass transition has to occur at $\theta_g = \theta_2$ because there is no more enthalpy relaxation at $T_g = T_2$ [31-33]. The existence of the time-lag $\tau^{ns}$, strongly varying with the temperature, has for consequence that the glass transition can occur in a wide window of temperatures.

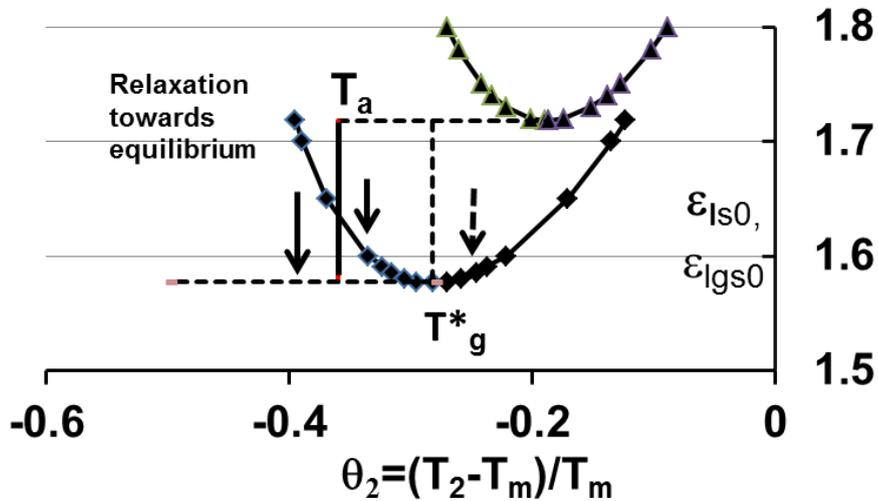

**Figure 2**. **Minimizing the energy saving coefficients.** The energy saving coefficients $\varepsilon_{ls0}$ (triangles) and $\varepsilon_{lgs0}$ (diamonds) of $Pd_{43}Ni_{10}Cu_{27}P_{20}$ have been calculated as a function of the homogeneous nucleation temperature $T_2$ using $T_{0m} = 430$ K, $T_{0g} = 365$ K, $T_m = 802$ K and (25). Note the minimum values of $\varepsilon_{ls0}$ and $\varepsilon_{lg0}$ given by (27) and (28) for "a" = 1. The transition at the temperature $T^*_g$ of the minimum transforms $\varepsilon_{ls0}$ into $\varepsilon_{lg0}$ and $T_{0m}$ into $T_{0g}$ during the relaxation time. A quench down to $T_a$ also leads to a transformation of $\varepsilon_{ls0}$ into $\varepsilon_{lg0}$ due to the existence of a minimum relaxation time between the two liquid states of about 100 s.

Slow physical vapor deposition rates produce vitreous samples having a new local packing arrangement as compared to that of bulk samples below $T_g$ and being at thermodynamic equilibrium. Indomethacin has been deposited on substrates cooled around the Kauzmann temperature in a highly stable vitreous state. This phenomenon of ultra-stable glass formation could be induced by "a first-order transition separating the normally observed high temperature liquid from a new low-temperature equilibrium supercooled liquid" [44]. The liquid-to-glass transformation is obtained at $T_g$ after 30,000 s instead of 100 s and accompanied by a strong change of local packing arrangement, as observed by wide-angle X-ray scattering measurements. These observations are compatible with the existence of a true liquid-to-glass transition which is produced at equilibrium by an endothermic latent heat during heating at $T_g$ after an exothermic latent heat at $T_K$. The progressive increase of the enthalpy saving cannot continue below the Kauzmann temperature [25]. The enthalpy excess of the undercooled liquid can only relax in a window of temperatures extending from $T_g$ to $T_K$. A saturation of the relaxed enthalpy has already been observed at $T_K$ [52].

The existence up to $T_g$ of highly-stable glasses, when they are prepared at equilibrium by physical vapor deposition, has the result that the current transformation at $T_g$, observed up to now without latent heat, is a liquid-to-liquid transition occurring in a pseudo-equilibrium state without enthalpy relaxation at a well-defined temperature $T_g$. The isothermally-relaxed enthalpy decreases when the annealing temperature increases up to $T_g$ and the enthalpy recovery measured at $T_g$ has to be equal to zero. This pseudo-equilibrium liquid state is not the equilibrium vitreous state and is obtained after an isothermal relaxation at a temperature $T_a$ higher than $T_K$ and lower than $T_g$. Equation (29) shows that the ultimate relaxed enthalpy to attain the new undercooled liquid state at $T_K$ is equal to $-\Delta\varepsilon_{lg0} \times \Delta H_m = -(\varepsilon_{ls0} - \varepsilon_{lgs0}) \times \Delta H_m$ and is recovered and included in the endothermic latent $L^+$ at $T_g$. A complementary exothermic enthalpy has to be relaxed at $T_K$ after a long steady-state nucleation time to give rise to the ultra-stable vitreous state. This exothermic latent heat produced at $T_K$ has to be equal to the endothermic enthalpy at $T_g$ given in (29) due to the thermodynamic character of the liquid-to-stable-glass transition.

**6. Scaling laws**

Equations (28-30) are scaling laws determining the new energy saving coefficient $\varepsilon_{lgs0}$ occurring at $\theta_2 = \theta_g$ and the temperatures $T_{0g}$ (and $\theta_{0g}$) and $T_{0m}$ (and $\theta_{0m}$) of fragile liquids from the knowledge of the thermodynamic transition temperature, which is defined as the disappearance temperature of the relaxed enthalpy [31,33]. The homogeneous nucleation temperature $T_1$ (or $\theta_1$) is determined as a function of $\varepsilon_{ls0}$ using (27) and follows a scaling law because, in many examples of fragile liquids, the viscosity is scaled by a VFT temperature $T_0 = 0.77\, T_g$ with $-0.35 < \theta_g < -0.2$ [41,42]. Scaling laws are easily obtained from (30-32), with a = 1 leading to $\varepsilon_{ls0} = \theta_g + 2$, $\Delta\varepsilon_{lg0} = -0.5 \times \theta_g$ and to a latent heat during heating equal to $-0.25 \times \theta_g$ due to the transformation of the stable glass in an undercooled liquid state:

$$\varepsilon_{ls0} = \varepsilon_{ls}(\theta = 0) = a \times \theta_g + 2, \tag{30}$$

$$\Delta\varepsilon_{lg0} = -0.5 \times \theta_g + (a-1)\theta_g, \tag{31}$$

$$\Delta H(\theta) = \Delta\varepsilon_{lg} \times \Delta H_m = \left[-0.5 \times \theta_g + (a-1)\theta_g + \frac{9\theta^2}{4\theta_g}\left(\frac{1}{a} - \frac{2}{3}\right)\right] \times \Delta H_m \longrightarrow L^- = (1-a)\theta_g \Delta H_m \tag{32}$$

The equilibrium enthalpies at $\theta = \theta_K$ and $\theta_g$ contain in (32) a contribution independent of temperature and a contribution proportional to $\theta^2$. This equilibrium enthalpy has a minimum value at $\theta_K$ when the undercooled liquid is the most unstable. It contains the maximum relaxed enthalpy at $T_K$ which is equal to $-0.5 \times \theta_g \times \Delta H_m$ and recovered at $T_g$, the endothermic latent heat $(a-1)\theta_g \times \Delta H_m$ delivered at $\theta_g$ when "a" is less than 1 and a contribution depending on $\theta^2$. The latent heat $L^+(\theta_g)$ becomes endothermic by heating at $\theta_g$ and equal to $\Delta H(\theta_g)$. The latent heat $L^-(\theta_g)$ obtained by cooling is equal to zero for a = 1 and only exists when "a" is less than 1 because the liquid is more fragile than a liquid with a = 1 with its higher value of $T_{0m}$. Measurements of $As_2Se_3$ and $As_2S_3$ showing the existence of a "glass-formation heat" obtained by cooling at $T_g$ demonstrates the existence of $L^-$ [52]. The total endothermic latent heat produced by heating at $T_g$ would be equal to $L^+$ for a << 1 and to the enthalpy recovery $L^+ = -0.25 \times \theta_g \times \Delta H_m$ for a = 1 if the stable glass phase has been previously formed by a long ageing at $T_K$. There is no latent heat associated with a stable-glass formation at $T_K$ in a strong liquid. It does not exist because $\Delta \varepsilon_{lg}$ is equal to zero at $T_g$ while a relaxed enthalpy equal to $\Delta \varepsilon_{lg0} \times \Delta H_m$ is still produced at $T_K$ when the transient relaxation time $\tau^{ns}$ is not evolved after quenching. The transformation in a stable glass is realized when this time-lag is evolved.

## 7. Specific heat jumps at the fragile-to-fragile, strong-to-strong, strong-to-fragile liquid transitions at $T_g$

### 7.1. Fragile-to-fragile liquid transition

A fragile-to-fragile liquid transition induces a new liquid state. The enthalpy derivative $[d(\Delta H)/dT]$ calculated using (29) or (32) is equal to the specific heat difference $\Delta C_p(T)$ between a quenched fragile undercooled liquid and its new equilibrium state given by (33), as already shown in 2008, without knowing at that time the exact values of thermodynamic parameters [26]:

$$\Delta C_p(T) = \Delta H_m \left[ \frac{d(\varepsilon_{ls} - \varepsilon_{lgs})}{dT} \right] = 2\theta \times \Delta S_m \left[ \frac{\varepsilon_{lgs0}}{\theta_{0g}^2} - \frac{\varepsilon_{ls0}}{\theta_{0g}^2} \right]. \tag{33}$$

The equilibrium specific heat jump below $T_g$ is defined by (34):

$$\Delta C_p(T) = 2 \frac{T - T_m}{T_g - T_m} \Delta S_m \left[ \frac{9}{4a} - \frac{9}{6} \right], \tag{34}$$

When a = 1, the scaling law is obeyed and the jump at $T_g$ is equal to (35):

$$\Delta C_p(T_g) = 1.5 \times \Delta S_m. \tag{35}$$

Equation (35) is respected in many glass-forming melts as shown by [37](p. 48) with many $\Delta C_p(T_g)$ jumps extending Wunderlich's previous finding [53]. The experimental values of $\Delta C_p(T_g)$ are used in part 8 to calculate the number "a" when (35) is not respected, including the experimental uncertainties on the measurements of various thermodynamic parameters.

### 7.2. Strong-to-fragile liquid transition

A strong-to-fragile liquid transition also induces a new liquid state. This phenomenon occurs when the fragile-to-fragile liquid transition temperature is expected to be a little lower than $0.5 \times T_m$ ($\theta_g < -0.5$). The relation (28) is obeyed because the liquid is fragile above $T_g$ with $\theta_{0m} > -2/3$ and becomes strong below $T_g$ with $\theta_{0g} \leq -2/3$. Equations (28, 36-38) are still used in part 7.3 to calculate ($\varepsilon_{ls0}$-$\varepsilon_{lgs0}$), $\theta_{0m}$, $T_{0m}$, $\Delta C_p(T_g)$, $\varepsilon_{ls0}$, $\varepsilon_{lgs0}$ and $\theta_{0g}$, which is now respecting the inequality $-1 < \theta_{0g} < -2/3$ (and $0 < T_{0g} < T_m/3$). The $Fe_{50}Co_{50}$ glass-forming melt transition is a good example of this phenomenon [54]. The thermodynamic parameters of several glasses undergoing this type of transition are presented in Table 1-1. The quantities $T_{0m}$, $T_{0g}$, $\Delta \varepsilon = (\varepsilon_{ls0}$-$\varepsilon_{lgs0})$ given in Table 1-1 are calculated only knowing the experimental values of $\Delta C_{plg}(T_g)$, $\Delta H_m$, $T_g$ and $T_m$. All $T_{0m}$ values of these fragile liquids are larger than $T_m/3$, whereas their $T_{0g}$ values are smaller than $T_m/3$. The temperatures $T_{0m}$ are not very different from the extrapolated VFT temperatures: 334/335, 768/768, 650/716, 417/372, 217/241 [ [55]].

**Table 1.1. Strong-to-fragile liquid transformations at the vitreous transition**.

**Table 1.2. Strong-to-strong liquid transformations at the vitreous transition.**
Heat capacity units are joules per gram.atom K. Fusion heat $\Delta H_m$ are given in kilojoules per gram.atom. The energy saving coefficient $\Delta\varepsilon_0$ is equal to the difference ($\varepsilon_{ls0}$-$\varepsilon_{lgs0}$).

| Materials | $T_m$ | $T_g$ | $\theta_g$ | $\Delta H_m$ | $\Delta C_p$ | $\Delta\varepsilon_0$ | $\varepsilon_{ls0}$ | $\varepsilon_{lgs0}$ | $T_{0m}$ | $T_{0g}$ | Ref. |
|---|---|---|---|---|---|---|---|---|---|---|---|
| **1-1 Fragile-to-strong** | | | | | | | | | | | |
| CaAl$_2$Si$_2$O$_8$ | 1830 | 1160 | -0.366 | 10.23 | 7.29 | 0.239 | 1.48 | 1.241 | 760 | 552 | [55,60] |
| As$_2$Te$_{3.13}$ | 649 | 391 | -0.397 | 11.16 | 16.9 | 0.195 | 1.408 | 1.213 | 254 | 203 | [61,62] |
| CaMgSi$_2$O$_6$ | 1670 | 1005 | -0.398 | 13.77 | 7.8 | 0.189 | 1.4 | 1.211 | 650 | 520 | [55,63,67] |
| Zr$_{46.75}$Ti$_{8.25}$Cu$_{7.5}$Ni$_{10}$Be$_{27.5}$ | 1070 | 640 | -0.402 | 10.63 | 9.9 | 0.200 | 1.401 | 1.201 | 417 | 331 | [64,65] |
| Au$_{77}$Ge$_{13.6}$Si$_{9.4}$ | 625 | 294 | -0.530 | 10.60 | 23.6 | 0.367 | 1.2 | 0.833 | 217 | 160 | [66] |
| **1-2 Strong-to-strong** | | | | | | | | | | | |
| NaAlSi$_3$O$_8$ | 1373 | 1096 | -0.202 | 4.83 | 2.1 | 0.0588 | 1.513 | 1.454 | 381 | 0 | [55,67] |
| SiO$_2$ | 1996 | 1473 | -0.262 | 2.97 | 1.0 | 0.0879 | 1.391 | 1.303 | 531 | 0 | [55,68] |
| BeF$_2$ | 825 | 590 | -0.285 | 1.59 | 1.0 | 0.0741 | 1.321 | 1.247 | 180 | 0 | [56, 69,70] |
| GeO$_2$ | 1358 | 820 | -0.396 | 5.57 | 1.5 | 0.0717 | 1.034 | 0.963 | 199 | 0 | [68,70,71] |

*7.3. Strong-to-strong liquid transition*

A strong-to-strong liquid transition also induces a new stronger liquid state assuming that $\Delta\varepsilon_{lg}$ in (31) is equal to zero at $T_g$ in the absence of minimum values of $\varepsilon_{ls0}$ and $\varepsilon_{lg0}$ when $T_{0m}$ is less than $T_m/3$. In this case, the specific heat jump becomes smaller than (35) and equal to (36):

$$\Delta C_p(T_g) = 2 \times \Delta S_m \frac{\varepsilon_{ls0} - \varepsilon_{lgs0}}{\theta_g}. \tag{36}$$

Equation (26) applied at the vitreous transition is used to determine $\varepsilon_{lgs0}$ with (37):

$$\varepsilon_{lgs0} = \frac{3\theta_g + 2}{1 - \theta_g^2 / \theta_{0g}^2}. \tag{37}$$

The specific heat jump is much smaller in strong glasses because the glass viscosity has to follow an Arrhenius law with $T_{0g} = 0$ K and $\theta_{0g} = -1$ instead of a negative value of $T_{0g}$ which would increase $\Delta C_p$; the stronger the glass-forming melt, the smaller the ($\varepsilon_{ls0}$-$\varepsilon_{lgs0}$) value. Equations (36,38) are used to calculate ($\varepsilon_{ls0}$-$\varepsilon_{lgs0}$), $\theta_{om}$, $T_{0m}$, $\Delta C_p(T_g)$, $\varepsilon_{ls0}$ and $\varepsilon_{lgs0}$:

$$\frac{\varepsilon_{ls0}}{\theta_{0m}^2} = \frac{\varepsilon_{lgs0}}{\theta_{0g}^2} + \frac{\varepsilon_{ls0} - \varepsilon_{lgs0}}{\theta_g^2}. \tag{38}$$

The transformation parameters of strong liquids are given in Table 1-2. The temperatures $T_{0g}$ are chosen equal to 0 K and $T_{0m}$ equal to the VFT temperatures [55,56]. The known values $\Delta H_m$, $T_g$ and $T_m$ are used to calculate the specific heat jump $\Delta C_p(T_g)$, the energy saving coefficients $\varepsilon_{ls0}$, $\varepsilon_{lg0}$ above and below $T_g$ and their difference. The jumps $\Delta C_p(T_g)$ per g.atom are very small compared to a crystal specific heat equal to 25 J/at.g.K. The calculated and experimental values are 2.1 and 2.05, 1 and 2.6, 1 and 0, 1.5 and 2.09 respectively. They are in good agreement considering a measurement uncertainty of about 0.5 to 1 J/at.g.K.

## 8. Specific heat jumps from metallic and non-metallic glasses to undercooled liquids at the vitreous transition

The specific heat differences $\Delta C_{plx}(T_g)$ between some fragile metallic liquids and crystals are indicated in column 8 of Table 2. The specific heat jumps $\Delta C_{plg}(T_g)$ at the fragile-to-fragile liquid transition given in Table 2 and Table 3 with a = 1 are equal to $1.5 \times \Delta S_m$ as predicted by (35) and in agreement with other reports, within the measurement uncertainties of specific heat, fusion heat and melting temperature of all liquids. It has been recently found that the jumps $\Delta C_{plg}$ of many metallic glasses are equal to 13.7 ± 2 J/K/at.g [57]. Their fusion entropy is expected to be equal to 9.13 ± 1.3 J/K/g.at applying (35), as already observed in many metallic liquid elements [46]. Values of $\Delta C_{plg}(T_g)$ of materials N°3 and N°6 in Table 2 are deduced from the slope of the maximum relaxed enthalpy versus temperature [31].

Liquids with a << 1 in Tables 2 and 3 have a larger specific heat jump. The "a" values in Figure 3 are calculated at $T_g$ with (34) using the experimental values of $\Delta C_{plg}(T_g)$ and represented as a function of $T_g/T_m$. The transition temperatures $T_g$, which are used in all tables to calculate thermodynamic parameters, are close to the thermodynamic transition temperatures where the relaxed enthalpy is equal to zero [31,33]. This approximation has a weak influence on them. Values of "a" larger than 1 are used to define in Figure 3 an experimental uncertainty of ± 6.5% and of ± 13% on the specific heat jump in this model. The "a" values equal to 1 in Table 2 and Table 3 correspond to this uncertainty. The fusion enthalpies of $ZnCl_2$ N° 50 and $B_2O_3$ N°51 have been reduced to respect (35) because of the existence of crystallographic instabilities under pressure and then under Laplace pressure and of hidden polymorphs [58,59]. There are 36 glass-forming melts among 49 following the scaling law (35), with the 13 others following (34) with values of "a" smaller than 1.

**Table 2.** **Specific heat jumps at the vitreous transition of metallic glass-forming melts**. The units are Kelvin and Joule/at.g/K.

| N° | Materials | $T_g$ | $\theta_g$ | $\varepsilon_{lg0}$ | a | $\varepsilon_{ls0}$ | $\Delta C_{plx}$ | $\Delta C_{plg}$ | $1.5\Delta S_m$ | $T_{0m}$ | $T_{0g}$ | Ref. |
|---|---|---|---|---|---|---|---|---|---|---|---|---|
| 1 | $Pd_{40}Ni_{10}Cu_{30}P_{20}$ | 578 | -0.276 | 1.586 | 1 | 1.724 | 19.9 | 12.8 | 12.8 | 431 | 367 | [72,73] |
| 2 | $Pd_{43}Ni_{10}Cu_{27}P_{20}$ | 576 | -0.282 | 1.577 | 1 | 1.718 | 20.2 | 13 | 13.1 | 430 | 365 | [72,74,75] |
| 3 | $Zr_{44}Ti_{11}Ni_{10}Cu_{10}Be_{25}$ | 620 | -0.327 | 1.510 | 0.91 | 1.703 |  | 19.6 | 15.1 | 484 | 393 | [76] |
| 4 | $Zr_{41.2}Ti_{13.8}Cu_{12.5}Ni_{10}Be_{22.5}$ | 625 | -0.333 | 1.501 | 0.833 | 1.723 | 22.5 | 21 | 13.1 | 505 | 396 | [36,77,78] |
| 5 | $Pd_{40}Ni_{40}P_{20}$ | 582 | -0.342 | 1.488 | 1 | 1.658 | 20.9 | 15.9 | 15.9 | 440 | 369 | [72,80,81] |
| 6 | $Ti_{40}Zr_{25}Ni_8Cu_9Be_{18}$ | 600 | -0.369 | 1.447 | 1 | 1.631 |  | 4.4 | 4.4 | 459 | 383 | [82] |

| | | | | | | | | | | | |
|---|---|---|---|---|---|---|---|---|---|---|---|
| 7 | Pt$_{57.3}$Cu$_{14.6}$Ni$_{5.3}$P$_{22.8}$ | 509 | -0.379 | 1.431 | 1 | 1.621 | 24.2 | 20.8 | 20.9 | 391 | 327 | [83,84] |
| 8 | Zr$_{52.5}$Al$_{10}$Ni$_{14.6}$Cu$_{17.9}$Ti$_5$ | 675 | -0.381 | 1.428 | 0.8 | 1.695 | 19.8 | | 11.3 | 568 | 434 | [85] |
| 9 | Zr46Cu46Al8 | 715 | -0.385 | 1.422 | 1 | 1.615 | 15 | 11.3 | 10.4 | 552 | 460 | (86) |
| 10 | Zr$_{57}$Al$_{10}$Ni$_{12.6}$Cu$_{15.4}$Nb$_5$ | 682 | -0.388 | 1.417 | 0.838 | 1.675 | 20 | | 12.6 | 566 | 440 | [85] |
| 11 | Zr$_{58.5}$Cu$_{15.6}$Ni$_{12.8}$Al$_{10.3}$Nb$_{2.8}$ | 675 | -0.392 | 1.412 | 1 | 1.608 | 14.5 | 12.5 | 11.8 | 523 | 436 | [83,87] |
| 12 | Pr$_{55}$Ni$_{25}$Al$_{20}$ | 494 | -0.395 | 1.407 | 0.785 | 1.690 | 32 | | 17.0 | 423 | 319 | [88] |
| 13 | Pd$_{77.5}$Cu$_6$Si$_{16.5}$ | 637 | -0.397 | 1.405 | 1 | 1.603 | 14.5 | 12.1 | 12.1 | 495 | 412 | [80,89,90] |
| 14 | Zr$_{45}$Cu$_{39.3}$Al$_7$Ag$_{8.7}$ | 691 | -0.398 | 1.403 | 1 | 1.602 | 14.3 | 11.2 | 10.4 | 537 | 448 | [86,91] |
| 15 | Cu$_{47}$Ti$_{34}$Zr$_{11}$Ni$_8$ | 673 | -0.403 | 1.395 | 1 | 1.597 | 14.5 | | 15.0 | 525 | 437 | [85,92] |
| 16 | Mg$_{65}$Cu$_{25}$Y$_{10}$ | 428 | -0.421 | 1.369 | 1 | 1.579 | 16 | | 17.6 | 337 | 281 | [93] |
| 17 | Zr$_{65}$Cu$_{17.5}$Ni$_{10}$Al$_{7.5}$ | 657 | -0.426 | 1.361 | 1 | 1.574 | 14 | | 13.5 | 520 | 433 | [83,94,98] |
| 18 | La$_{55}$Al$_{25}$Ni$_5$Cu$_{10}$Co$_5$ | 466 | -0.433 | 1.350 | 1 | 1.567 | 16.7 | 11.8 | 11.1 | 370 | 309 | [96,97] |
| 19 | Zr$_{65}$Cu$_{27.5}$Al$_{7.5}$ | 666 | -0.436 | 1.347 | 1 | 1.564 | 15 | | 16.3 | 531 | 442 | [95,98] |
| 20 | La$_{55}$Al$_{25}$Ni$_{10}$Cu$_{10}$ | 467 | -0.440 | 1.340 | 1 | 1.560 | 15 | | 12.3 | 374 | 311 | [96,97] |
| 21 | La$_{55}$Al$_{25}$Ni$_{15}$Cu$_5$ | 472 | -0.476 | 1.287 | 1 | 1.524 | 14.9 | | 12.5 | 389 | 325 | [96,97] |
| 22 | La$_{55}$Al$_{25}$Ni$_5$Cu$_{15}$ | 459 | -0.477 | 1.284 | 1 | 1.523 | 13.8 | | 12.3 | 379 | 317 | [96,97] |
| 23 | La$_{55}$Al$_{25}$Ni$_{20}$ | 491 | -0.478 | 1.283 | 1 | 1.522 | 13.5 | | 11.9 | 406 | 339 | [96,97] |

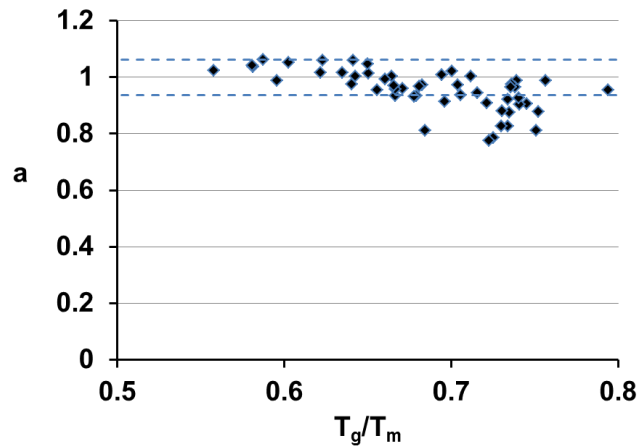

**Figure 3. The number "a" versus $T_g/T_m$.** The number "a" of fragile non-polymeric glass-forming liquids defined in (30) at $T_g$ is plotted as a function of $T_g/T_m$

**Table 3**. A collection of specific heat jumps measured in fragile liquids and selected by Wang, Angell and Richert [ [70]] are compared to $1.5 \times \Delta S_m$. The units are Kelvin and J/mole/K. The fusion entropy of N°17 has been changed using new measurements [99]. The $ZnCl_2$ and $B_2O_3$ fusion entropies are decreased to respect $\Delta C_{plg} = 1.5 \times \Delta S_m$ because of the existence of crystallographic instabilities under pressure and hidden polymorphs which are also acting under Laplace pressure [ [58], [59]].

| N° | Materials | $T_g$ | $\theta_g$ | $\varepsilon_{lg0}$ | a | $\varepsilon_{ls0}$ | $\Delta C_{plg}$ | $1.5\Delta S_m$ | $T_{0g}$ | $T_{0m}$ | Ref. |
|---|---|---|---|---|---|---|---|---|---|---|---|
| 1 | β-D-fructose | 286 | -0.243 | 1.635 | 1 | 1.757 | 133 | 128.7 | 183 | 213 | [100,101] |
| 2 | o-Terphenyl | 240 | -0.248 | 1.628 | 0.875 | 1.763 | 112 | 78.9 | 157 | 187 | [102,103] |
| 3 | m-Toluidine | 187 | -0.249 | 1.627 | 0.811 | 1.798 | 90 | 53.0 | 120 | 149 | [104,105] |
| 4 | Flopropione | 335 | -0.259 | 1.612 | 0.904 | 1.766 | 127.5 | 96.6 | 214 | 258 | [70,106,107] |
| 5 | Maltitol | 311 | -0.260 | 1.611 | 0.926 | 1.760 | 243.6 | 196.4 | 198 | 238 | [108,109] |
| 6 | Probucol | 295 | -0.261 | 1.609 | 1 | 1.739 | 139.5 | 134.1 | 188 | 220 | [106,107,110] |
| 7 | Griseofulvin | 364 | -0.262 | 1.608 | 1 | 1.738 | 127 | 114.9 | 232 | 271 | [106,107] |
| 8 | Indomethacin | 318 | -0.264 | 1.604 | 1 | 1.736 | 147 | 136.8 | 203 | 237 | [70,111] |
| 9 | D-glucose | 309 | -0.264 | 1.604 | 1 | 1.736 | 128 | 115.7 | 197 | 230 | [70,100,101] |
| 10 | PMS | 167 | -0.265 | 1.603 | 0.875 | 1.768 | 138 | 96.7 | 106 | 130 | [70,112] |
| 11 | Sucrose | 345 | -0.266 | 1.601 | 0.827 | 1.780 | 215 | 132.1 | 220 | 274 | [70,100,110] |
| 12 | Glibenclamide | 331 | -0.266 | 1.601 | 0.922 | 1.755 | 222.3 | 177.4 | 211 | 254 | [106,107] |
| 13 | Propylene Carbonate | 160 | -0.270 | 1.595 | 0.879 | 1.763 | 75.4 | 53.4 | 101 | 124 | [113] |
| 14 | Sorbitol | 268 | -0.270 | 1.595 | 0.827 | 1.777 | 201 | 123.4 | 170 | 213 | [70,110] |
| 15 | Li-Acetate | 401 | -0.283 | 1.576 | 0.781 | 1.779 | 62.7 | 34.1 | 254 | 325 | [70] |
| 16 | Triphenylethene | 246 | -0.279 | 1.582 | 0.907 | 1.747 | 117 | 89.5 | 156 | 190 | [114] |
| 17 | $As_2Se_3$ | 462 | -0.283 | 1.576 | 0.753 | 1.787 | 72 | 36.3 | 293 | 380 | [52,99] |
| 18 | 1,3,5-tri-α-Naphtylbenzene | 340 | -0.284 | 1.574 | 1 | 1.716 | 124 | 105.2 | 216 | 254 | [70,115] |
| 19 | Phenobarbital | 319 | -0.286 | 1.570 | 1 | 1.714 | 106.8 | 93.6 | 202 | 238 | [106,107] |
| 20 | Isopropylbenzene | 125 | -0.294 | 1.558 | 1 | 1.706 | 74.6 | 62.2 | 79 | 93 | [116] |
| 21 | Hydro-chloro-thiazide | 385 | -0.296 | 1.556 | 1 | 1.704 | 92.3 | 85.0 | 244 | 288 | [106,107] |
| 22 | 3-Methylpentane | 77 | -0.299 | 1.551 | 1 | 1.701 | 68 | 72.3 | 49 | 58 | [117] |
| 23 | Salol | 220 | -0.304 | 1.544 | 0.912 | 1.723 | 118 | 91.6 | 139 | 170 | [113,114] |
| 24 | m-Cresol | 199 | -0.306 | 1.542 | 1 | 1.694 | 54 | 55.4 | 126 | 149 | [104,105] |
| 25 | $Ca(NO_3)_2$-$4H_2O$ | 217 | -0.315 | 1.527 | 0.812 | 1.744 | 250 | 147.5 | 137 | 176 | [118] |
| 26 | Xylitol | 244 | -0.317 | 1.524 | 1 | 1.683 | 155 | 142.7 | 154 | 183 | [119] |
| 27 | Phenolphthalein | 363 | -0.319 | 1.522 | 1 | 1.681 | 146 | 132.7 | 230 | 273 | [70] |
| 28 | 9-Bromo phenanthrene | 225 | -0.321 | 1.519 | 1 | 1.679 | 77 | 63.4 | 142 | 169 | [113] |
| 29 | Triphenyl phosphite | 200 | -0.322 | 1.517 | 1 | 1.678 | 155 | 127.1 | 127 | 150 | [70] |
| 30 | α-Phenil -cresol | 220 | -0.329 | 1.506 | 1 | 1.671 | 120 | 106.6 | 139 | 166 | [120] |
| 31 | $H_2SO_4$-$3H_2O$ | 158 | -0.333 | 1.500 | 1 | 1.667 | 186 | 153.3 | 100 | 119 | [121,122] |
| 32 | Diethylphthalate | 178 | -0.333 | 1.500 | 1 | 1.667 | 115 | 101.1 | 113 | 134 | [123] |
| 33 | m-Fluorotoluene | 123 | -0.334 | 1.499 | 1 | 1.666 | 74 | 67.7 | 78 | 92 | [69,104,105] |
| 34 | 2-methyl tetrahydrofuran | 91 | -0.336 | 1.496 | 1 | 1.664 | 72 | 72.8 | 58 | 69 | [124,125] |
| 35 | n-Butene | 58 | -0.339 | 1.491 | 1 | 1.661 | 69 | 67.7 | 37 | 44 | [126,127] |
| 36 | Toluene | 117 | -0.344 | 1.483 | 1 | 1.656 | 64 | 55.8 | 74 | 89 | [128] |
| 37 | Glycerol | 190 | -0.349 | 1.476 | 1 | 1.651 | 90 | 94.0 | 121 | 144 | [113] |
| 38 | 2-Methyl pentane | 78 | -0.350 | 1.475 | 1 | 1.650 | 68 | 78.3 | 50 | 59 | [50,129] |
| 39 | Ethylbenzene | 115 | -0.358 | 1.464 | 1 | 1.642 | 76 | 76.8 | 73 | 88 | [128] |
| 40 | n-Propanol | 96 | -0.359 | 1.462 | 1 | 1.641 | 45 | 54.0 | 61 | 73 | [130] |

| 41 | **3-Bromopentane** | 106 | -0.365 | 1.452 | 1 | 1.635 | 72 | 75.4 | 68 | 81 | [124] |
| --- | --- | --- | --- | --- | --- | --- | --- | --- | --- | --- | --- |
| 42 | **2-Methyl-1-propanol** | 107 | -0.377 | 1.435 | 1 | 1.623 | 46 | 55.2 | 69 | 82 | [131] |
| 43 | **Selenium** | 309 | -0.378 | 1.433 | 1 | 1.622 | 14.4 | 15.1 | 198 | 237 | [103,132] |
| 44 | **Butyronitrile** | 97 | -0.398 | 1.404 | 1 | 1.602 | 40 | 46.8 | 63 | 75 | [133] |
| 45 | **Cis-/trans-Decalin** | 137 | -0.404 | 1.394 | 1 | 1.596 | 64 | 61.5 | 89 | 107 | [113] |
| 46 | **Ethanol** | 94 | -0.413 | 1.381 | 1 | 1.588 | 38 | 46.2 | 61 | 74 | [69,70,134-136] |
| 47 | **Methanol** | 100 | -0.419 | 1.372 | 1 | 1.581 | 30 | 33.6 | 66 | 79 | [50,136,137] |
| 48 | **Ethylene glycol** | 151 | -0.419 | 1.371 | 1 | 1.581 | 60 | 68.4 | 99 | 119 | [138] |
| 49 | **m-Xylene** | 126 | -0.442 | 1.337 | 1 | 1.558 | 72 | 77.1 | 84 | 101 | [104,105] |
| 50 | **ZnCl$_2$** | 378 | -0.359 | 1.461 | 1 | 1.641 | 17 | 17.0 | 241 | 288 | [58,139] |
| 51 | **B$_2$O$_3$** | 536 | -0.259 | 1.612 | 1 | 1.741 | 40 | 40 | 342 | 400 | [59,140,141] |

## 9. Enthalpy thermal cycles expected in some liquids and determination of the Kauzmann temperature

The ultimate relaxed enthalpy is the maximum relaxed enthalpy occurring at equilibrium at the Kauzmann temperature. The total equilibrium enthalpy changes at $T_g$ equal to the latent heat $L^+$ and $L^-$ as defined by (32) are given in Tables 4 and 5. The relaxed enthalpy of As$_2$Se$_3$ is saturated at the Kauzmann temperature, equal to 6.4 J/g corresponding to 2.48 kJ/mole and is approximately equal to that given by the scaling law $0.5 \times \theta_g \times \Delta H_m = -2.21$ kJ/mole [52]. This glass-forming melt does not follow the scaling law above $T_g$, as shown in Table 3 N°17. Its thermodynamic parameters given in Table 3 and 5 are $T_m = 645$ K, $T_{0m} = 380$ K, $T_{0g} = 293$ K, $\theta_g = -0.283$, a = 0.753, $\Delta H_m = 15.6$ kJ/mole, $L^+ = 0.21 \times \Delta H_m = 3.28$ kJ/mole and $L^- = 0.07 \times \Delta H_m = 1.09$ kJ/mole. An endothermic enthalpy $L^+$ of 3.5 kJ/mole (9.04 J/g) corresponding to $\Delta \varepsilon_{lg} = 0.224$ has been measured at $T_g$ by transforming the liquid state below $T_g$ into the undercooled liquid state above $T_g$ after a long ageing time near $T_K$ equal to 166 hours. The application of (32) using a = 0.753 leads to a latent heat $L^- = 0.07 \times \Delta H_m = 1.09$ kJ/mole and $L^+ = 0.210 \times \Delta H_m = 3.28$ kJ/mole which are approximately equal to experimental values of 1.07 kJ/mole and 3.5 kJ/mole respectively [52](Table 1). These experimental results confirm that the ultimate relaxed enthalpy always respects the scaling law whatever the number "a" may be and that an exothermic latent heat can be observed at $T_g$ while cooling a glass-forming melt through the vitreous transition when the number "a" is much smaller than 1 [33].

The ultimate enthalpy recovery of butyronitrile has also been studied by measuring the relaxed enthalpy after vapor deposition at 40 K, far below $T_K$. It is equal to 1.3 kJ/mole and larger than $0.5 \times \theta_g \times \Delta H_m = 1$ kJ/mole and smaller than the equilibrium enthalpy of the stable glass phase equal to 1.5 kJ/mole expected at $T_K$ including the latent heat $0.25 \times \theta_g \times \Delta H_m$ associated with a stable glass formation when a = 1 [133]. The out-of-equilibrium entropy of the undercooled liquid remains larger than that of crystals. In these conditions, the ultimate relaxed enthalpy $\Delta H_r$ equal to $0.5 \times \theta_g$ in all liquids can be used to calculate $T_K$ considering that all thermodynamic properties are obeying scaling laws below $T_g$. The maximum relaxed enthalpy decreases with temperature. Its derivative $dH_r/dT$ is equal to a specific heat difference $\Delta C_{plg}(T)$, being nearly constant below $T_g$ and nearly equal to $\Delta C_{plg}(T_g)$.

The Kauzmann temperature $T_K$ is calculated using the mean theoretical specific heat below $T_g$ deduced from (33) and imposing the ultimate enthalpy recovery to be given by (39). The $\theta_K$ is finally given by (40) with the number "a" determined by imposing the theoretical specific heat jump to be equal to the experimental one:

$$\overline{\Delta C_{p\lg}(T \leq T_g)} \times (\Delta T_K) = -0.5 \times \theta_g \times \Delta H_m, \quad (39)$$

$$\theta_K^2 = (\frac{T_K - T_m}{T_m})^2 = \theta_g^2 - 0.5 \times \theta_g \times (\frac{\varepsilon_{ls0}}{\theta_{0m}^2} - \frac{\varepsilon_{\lg s0}}{\theta_{0g}^2})^{-1} = \theta_g^2 + 0.5 \times \theta_g^2 \times \frac{4a}{(9-6a)}. \quad (40)$$

In many cases for which a = 1, the scaling law $\theta_K^2 = \frac{5}{3}\theta_g^2$ is respected.

The equilibrium enthalpies divided by the fusion heat are equal to $-\Delta\varepsilon_{lg}$ and represented in Figures 4, 5 and 6 for $Pd_{43}Ni_{10}Cu_{27}P_{20}$ BMG N° 2 (a = 1), indomethacin Glass N°8 (a = 1) and $As_2Se_3$ Glass N°17 (a = 0.776) as a function of temperature below $T_g$. There is no latent heat expected by cooling with a = 1 because the temperature $T_g$ only corresponds to a liquid-liquid transition which attains a pseudo-equilibrium after relaxation. On the contrary, a partial latent heat equal to $L^-$ is expected for "a" << 1. The out-of-equilibrium liquid quenched below $T_g$ from high temperatures to $T_K$ can be totally transformed into a stable glass phase after ageing at $T_K$ when the nucleation time is minimum. The ultimate enthalpy has to be relaxed to attain first the pseudo-equilibrium liquid state and, after a much longer time, the stable glass phase producing an exothermic latent heat. An endothermic latent heat $L^+$ is needed to transform the stable glass phase into an undercooled liquid at $T_g$.

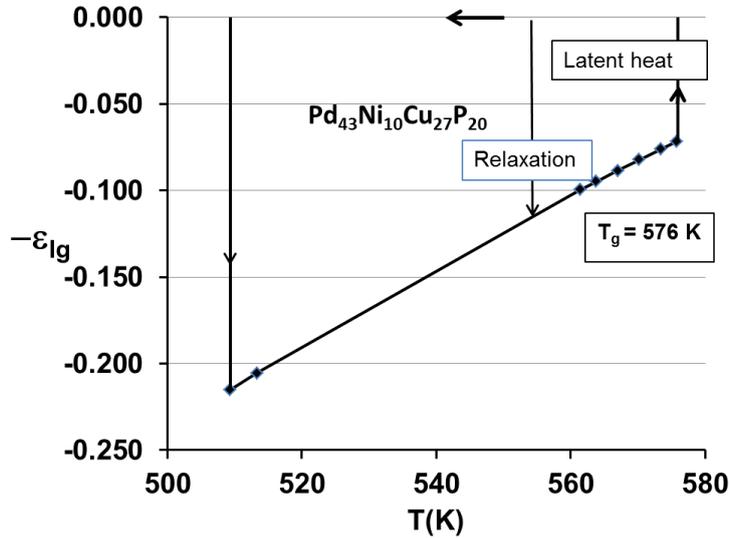

**Figure 4. The stable glass phase formation at $T_K$ in $Pd_{43}Ni_{10}Cu_{27}P_{20}$.** The enthalpy variation from the quenched to the equilibrium state of the stable glass phase obtained at $T_K$ divided by $\Delta H_m$ is represented by $-\Delta\varepsilon_{lg}$ as a function of temperature. The irreversible and endothermic latent heats at $T_g$ are equal to $-0.25 \times \theta_g \times \Delta H_m$ (a = 1).

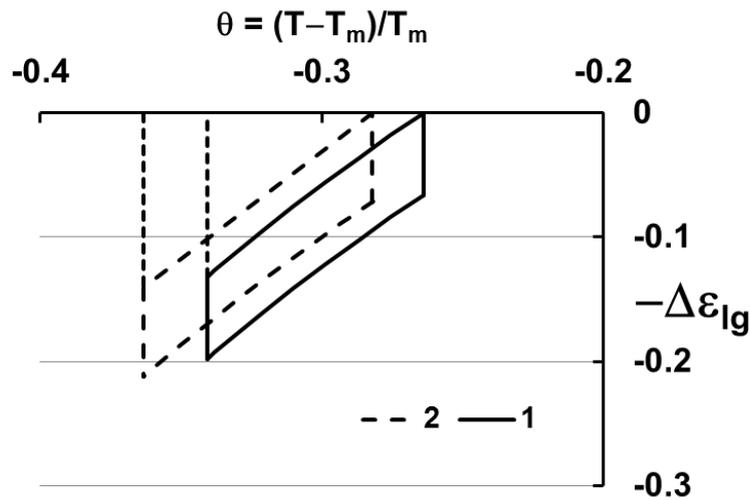

**Figure 5. The stable glass phase formation between $T_K$ and $T_g$ in $Pd_{43}Ni_{10}Cu_{27}P_{20}$ and indomethacin.** The enthalpy saving of indomethacin below the vitreous transition $T_g$ represented by the coefficient $-\Delta\varepsilon_{lg}$ is plotted versus temperature and compared to that of $Pd_{43}Cu_{23}Ni_{10}P_{20}$. There is no reversible latent heat in the liquid-to-liquid transition at $T_g$ even if a latent heat $L^+$ is expected during the heating of the stable glass phase. These two liquids undergo a phase transition at $T_g$ characterized by a change in the enthalpy slope at $T_g$ because "a" = 1. Pseudo-equilibrium enthalpies of undercooled melts are obtained during cooling after relaxation at the annealing temperature T. After a long ageing at $T_K$, the transitions to the stable glass states would be accompanied by an exothermic latent heat and the stable glass enthalpies would increase up to $T_g$. An irreversible endothermic latent heat equal to $-0.25\times\theta_g\times\Delta H_m$ is needed to return to the equilibrium undercooled liquid state above $T_g$.

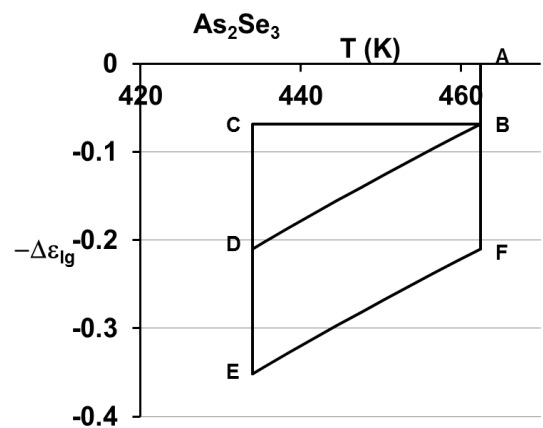

**Figure 6. The formation of stable glass phase in $As_2Se_3$.** The enthalpy variation divided by $\Delta H_m$ represented by $-\Delta\varepsilon_{lg}$ as a function of the temperature. A reversible latent heat occurs along AB and BA at $T_g$ during cooling and heating because "a" << 1. There is no more enthalpy change along BC when the undercooled liquid is rapidly cooled without being relaxed. The structural relaxation progressively transforms BC into BD and the undercooled liquid is transformed into a new liquid state, producing an exothermic relaxed enthalpy. After a long ageing along DE, the undercooled liquid is transformed into a stable glass phase. The glass phase enthalpy increases along EF by heating. An irreversible endothermic enthalpy is produced by heating along FB.

The temperature difference $\Delta T_K = (T_g - T_K)$ is compared in many of the examples given in Tables 4 and 5 to experimental values obtained by entropy extrapolations from temperatures above $T_g$ using specific heat laws measured from $T_g$ to $T_m$. These quantities are nearly equal for "a" = 1 in many cases. For a < 1, they cannot be equal, as shown in Tables 4 and 5, because the contribution of the latent heat $L^-$ to the available entropy is not subtracted in the extrapolation method.

In addition, the liquid specific heat excess below $T_g$ is nearly constant, as demonstrated by a maximum relaxed enthalpy linearly decreasing with temperature up to $T_g$ as shown in Figures 4, 5 and 6. The presence of a hidden freezing before relaxation at $T_g$ is demonstrated in part 11 and the existence of a complementary reduction of $\Delta T_K$ for a < 1 is explained. The frozen enthalpy values $\Delta H_g$ given in Tables 4 and 5 are equal to $(L^+ - 0.5 \times \theta_g \times \Delta H_m)$ and always smaller than the available enthalpy

$$\Delta H_m - \int_{Tg}^{Tm} \Delta C_{pls} dT$$

as already predicted [103]. The frozen entropy is also smaller than the available entropy

$$\Delta S_m - \int_{Tg}^{Tm} \frac{\Delta C_{pls}}{T} dT$$

because $|\Delta C_{plg}|$ is always smaller than the specific heat difference $|\Delta C_{plx}|$ between undercooled liquid and crystal.

**Table 4. Thermodynamics parameters of some bulk metallic glasses**: The equilibrium endothermic latent heats $L^+$ and $L^-$ divided by $\Delta H_m$, the difference $\Delta T_K = (T_g - T_K)$ and the frozen enthalpy $\Delta H_g/\Delta H_m = (-0.5 \times \theta_g + L^+)$ below $T_g$ divided by $\Delta H_m$ of some fragile metallic glass-forming melts. Fusion heat units are kilojoules per g.atom. The maximum value of the relaxed enthalpy is equal to $-0.5 \times \theta_g \times \Delta H_m$; $\Delta T_K = T_g - T_K$; $L^+$ and $L^-$ are equilibrium latent heats.

| N° | Materials | $T_m$ | $\Delta H_m$ | a | $\Delta H_g/\Delta H_m$ | $L^-/\Delta H_m$ | $L^+/\Delta H_m$ | $\Delta T_{Kcalc}$ | $\Delta T_{Kexp}$ | Ref |
|---|---|---|---|---|---|---|---|---|---|---|
| 1 | $Pd_{40}Ni_{10}Cu_{30}P_{20}$ | 798 | 6.82 | 1 | 0.207 | 0 | 0.069 | 64 | 71 | [72] |
| 2 | $Pd_{43}Ni_{10}Cu_{27}P_{20}$ | 802 | 7.01 | 1 | 0.211 | 0 | 0.070 | 66 | 79 | [72] |
| 3 | $Zr_{44}Ti_{11}Ni_{10}Cu_{10}Be_{25}$ | 921 | 9.30 | 0.91 | 0.275 | 0.029 | 0.125 | 69 | 75 | [33,76] |
| 4 | $Zr_{41.2}Ti_{13.8}Cu_{12.5}Ni_{10}Be_{22.5}$ | 937 | 8.20 | 0.833 | 0.305 | 0.056 | 0.178 | 59 | 65 | [77,78] |
| 5 | $Pd_{40}Ni_{40}P_{20}$ | 884 | 9.40 | 1 | 0.256 | 0 | 0.085 | 88 | 80 | [81] |
| 6 | $Ti_{40}Zr_{25}Ni_8Cu_9Be_{18}$ | 950 | 2.78 | 1 | 0.277 | 0 | 0.092 | 102 | | |
| 7 | $Pt_{57.3}Cu_{14.6}Ni_{5.3}P_{22.8}$ | 820 | 11.40 | 1 | 0.284 | 0 | 0.095 | 90 | 113 | [84] |
| 8 | $Zr_{52.5}Al_{10}Ni_{14.6}Cu_{17.9}Ti_5$ | 1091 | 8.20 | 0.8 | 0.362 | 0.076 | 0.234 | 73 | 37 | [85] |
| 9 | $Zr_{46}Cu_{46}Al_8$ | 1163 | 8.04 | 1 | 0.289 | 0 | 0.096 | 130 | 119 | [86] |
| 10 | $Zr_{57}Al_{10}Ni_{12.6}Cu_{15.4}Nb_5$ | 1115 | 9.40 | 0.838 | 0.354 | 0.063 | 0.203 | 83 | 18 | [85] |
| 11 | $Zr_{58.5}Cu_{15.6}Ni_{12.8}Al_{10.3}Nb_{2.8}$ | 1110 | 8.70 | 1 | 0.294 | 0 | 0.098 | 127 | 127 | [87] |
| 12 | $Pr_{55}Ni_{25}Al_{20}$ | 817 | 9.24 | 0.785 | 0.382 | 0.085 | 0.257 | 55 | 44 | [88] |
| 13 | $Pd_{77.5}Cu_6Si_{16.5}$ | 1056 | 8.55 | 1 | 0.298 | 0 | 0.099 | 122 | 73 | [80] |
| 14 | $Zr_{45}Cu_{39.3}Al_7Ag_{8.7}$ | 1148 | 7.93 | 1 | 0.299 | 0 | 0.100 | 133 | | [91] |
| 15 | $Cu_{47}Ti_{34}Zr_{11}Ni_8$ | 1128 | 11.30 | 1 | 0.303 | 0 | 0.101 | 132 | 136 | [85] |
| 16 | $Mg_{65}Cu_{25}Y_{10}$ | 739 | 8.65 | 1 | 0.316 | 0 | 0.105 | 90 | 103 | [93] |
| 17 | Zr$_{65}$Cu$_{17.5}$Ni$_{10}$Al$_{7.5}$ | 1145 | 10.30 | 1 | 0.320 | 0 | 0.107 | 142 | 136 | [98,142] |
| 18 | La$_{55}$Al$_{25}$Ni$_5$Cu$_{10}$Co$_5$ | 822 | 6.08 | 1 | 0.325 | 0 | 0.108 | 104 | 103 | [96] |
| 19 | Zr$_{65}$Cu$_{27.5}$Al$_{7.5}$ | 1180 | 12.80 | 1 | 0.327 | 0 | 0.109 | 150 | | |
| 20 | La$_{55}$Al$_{25}$Ni$_{10}$Cu$_{10}$ | 835 | 6.82 | 1 | 0.330 | 0 | 0.110 | 107 | 135 | [96] |
| 21 | La$_{55}$Al$_{25}$Ni$_{15}$Cu$_5$ | 900 | 7.49 | 1 | 0.357 | 0 | 0.119 | 125 | 154 | [96] |
| 22 | La$_{55}$Al$_{25}$Ni$_5$Cu$_{15}$ | 878 | 7.19 | 1 | 0.358 | 0 | 0.119 | 122 | 155 | [96] |
| 23 | La$_{55}$Al$_{25}$Ni$_{20}$ | 941 | 7.46 | 1 | 0.359 | 0 | 0.120 | 131 | 154 | [96] |

**Table 5. Thermodynamics parameters of some glasses**: The equilibrium endothermic latent heats $L^+$ and $L^-$ divided by $\Delta H_m$, the difference $\Delta T_K = (T_g - T_K)$ and the frozen enthalpy $\Delta H_g/\Delta H_m = (-0.5 \times \theta_g + L^+)$ between $T_K$ and $T_g$ divided by $\Delta H_m$ of some fragile metallic glass-forming melts. The relaxed ultimate enthalpy is equal to $-0.5 \times \theta_g \times \Delta H_m$. The fusion enthalpy of N°17 is changed using new measurements [52]. The ZnCl$_2$ and B$_2$O$_3$ fusion enthalpies are also changed to respect $\Delta C_{plg} = 1.5 \times \Delta S_m$ because there exists a crystallographic instability under pressure [58,59] and then under Laplace pressure.

| | Materials | $T_m$ (K) | $\Delta H_m$ (kJ/mol) | a | $\Delta\varepsilon_0$ | $L^+/\Delta H_m$ | $L^-/\Delta H_m$ | $\Delta H_g/\Delta H_m$ calc | $\Delta T_K$ calc | $\Delta T_K$ exp | Ref |
|---|---|---|---|---|---|---|---|---|---|---|---|
| 1 | **β-D-fructose** | 378 | 32.43 | 1 | 0.122 | 0.061 | 0 | 0.189 | 27 | 76 | [50] |
| 2 | **o-Terphenyl** | 329 | 17.2 | 0.875 | 0.170 | 0.121 | 0.034 | 0.251 | 19 | 40 | [103] |
| 3 | **m-Toluidine** | 249 | 8.8 | 0.811 | 0.171 | 0.146 | 0.047 | 0.273 | 11 | 36 | [50] |
| 4 | **Flopropione** | 452 | 29.1 | 0.904 | 0.154 | 0.102 | 0.025 | 0.232 | 27 | | |
| 5 | **Maltitol** | 420 | 55 | 0.926 | 0.149 | 0.092 | 0.019 | 0.222 | 26 | | |
| 6 | **Probucol** | 399 | 35.66 | 1 | 0.130 | 0.065 | 0.000 | 0.196 | 30 | | |
| 7 | **Griseofulvin** | 493 | 37.75 | 1 | 0.131 | 0.065 | 0.000 | 0.197 | 38 | | |
| 8 | **Indomethacin** | 432 | 39.4 | 1 | 0.132 | 0.066 | 0.000 | 0.198 | 33 | 70 | [111] |
| 9 | **D-glucose** | 420 | 32.4 | 1 | 0.132 | 0.066 | 0.000 | 0.198 | 32 | 38 | [50] |
| 10 | **PMS** | 227 | 14.65 | 0.875 | 0.165 | 0.118 | 0.033 | 0.251 | 13 | 30 | [50] |
| 11 | **Sucrose** | 470 | 41.4 | 0.827 | 0.179 | 0.146 | 0.046 | 0.279 | 23 | 62 | [50] |
| 12 | **Glibenclamide** | 451 | 53.35 | 0.922 | 0.154 | 0.096 | 0.021 | 0.230 | 29 | | |
| 13 | **Propylene Carbonate** | 218 | 7.77 | 0.879 | 0.167 | 0.118 | 0.033 | 0.253 | 13 | 33.7 | |
| 14 | **Sorbitol** | 367 | 30.2 | 0.827 | 0.182 | 0.148 | 0.047 | 0.286 | 19 | 19 | [110] |
| 15 | **Li-Acetate** | 559 | 12.7 | 0.781 | 0.203 | 0.187 | 0.062 | 0.327 | 26 | | |
| 16 | **Triphenylethene** | 341 | 20.35 | 0.907 | 0.165 | 0.108 | 0.026 | 0.248 | 22 | | |
| 17 | **As$_2$Se$_3$** | 645 | 15.6 | 0.753 | 0.205 | 0.225 | 0.070 | 0.366 | 28 | | [52,99] |
| 18 | **1,3,5-tri-α-Naphtylbenzene** | 475 | 33.3 | 1 | 0.142 | 0.071 | 0 | 0.214 | 39 | | |
| 19 | **Phenobarbital** | 447 | 27.9 | 1 | 0.143 | 0.072 | 0 | 0.217 | 37 | | |

| | | | | | | | | | | |
|---|---|---|---|---|---|---|---|---|---|---|
| 20 | **Isopropyl benzene** | 177 | 7.33 | 1 | 0.147 | 0.074 | 0 | 0.221 | 15 | | |
| 21 | **Hydro-chloro-thiazide** | 547 | 31 | 1 | 0.148 | 0.074 | 0 | 0.223 | 47 | | |
| 22 | **3-Methylpentane** | 110 | 53 | 1 | 0.150 | 0.075 | 0 | 0.226 | 10 | | |
| 23 | **Salol** | 316 | 19.3 | 0.912 | 0.179 | 0.115 | 0.027 | 0.267 | 22 | 53 | [50] |
| 24 | **m-Cresol** | 286 | 10.57 | 1 | 0.153 | 0.076 | 0 | 0.232 | 25 | | [105] |
| 25 | **Ca(NO$_3$)$_2$-4H$_2$O** | 317 | 31.17 | 0.812 | 0.217 | 0.184 | 0.059 | 0.342 | 18 | | [118] |
| 26 | **Xylitol** | 358 | 34 | 1 | 0.159 | 0.079 | 0 | 0.238 | 33 | | |
| 27 | **Phenolphthalein** | 533 | 47.15 | 1 | 0.159 | 0.080 | 0 | 0.240 | 49 | 53 | [50] |
| 28 | **9-Bromophenanthrene** | 331 | 14 | 1 | 0.160 | 0.080 | 0 | 0.241 | 31 | | |
| 29 | **Triphenyl phosphite** | 295 | 25 | 1 | 0.161 | 0.081 | 0 | 0.243 | 28 | | |
| 30 | **α-Phenil -cresol** | 328 | 23.3 | 1 | 0.165 | 0.082 | 0 | 0.248 | 31 | | |
| 31 | **H$_2$SO$_4$-3H$_2$O** | 237 | 24.22 | 1 | 0.167 | 0.083 | 0 | 0.250 | 23 | 23 | [50] |
| 32 | **Diethylphthalate** | 267 | 17.99 | 1 | 0.167 | 0.083 | 0 | 0.250 | 26 | | |
| 33 | **m-Fluorotoluene** | 184 | 8.3 | 1 | 0.167 | 0.084 | 0 | 0.251 | 18 | | |
| 34 | **2-methyl tetrahydrofuran** | 137 | 6.65 | 1 | 0.168 | 0.084 | 0 | 0.253 | 13 | | |
| 35 | **n-Butene** | 88 | 3.96 | 1 | 0.170 | 0.085 | 0 | 0.256 | 9 | 10 | [127] |
| 36 | **Toluene** | 178 | 6.64 | 1 | 0.172 | 0.086 | 0 | 0.260 | 18 | 21 | [104,105] |
| 37 | **Glycerol** | 292 | 18.3 | 1 | 0.175 | 0.087 | 0 | 0.262 | 30 | 55 | [50] |
| 38 | **2-Methylpentane** | 120 | 6.26 | 1 | 0.175 | 0.088 | 0 | 0.264 | 12 | 20 | [129] |
| 39 | **Ethylbenzene** | 179 | 9.17 | 1 | 0.179 | 0.089 | 0 | 0.268 | 19 | | |
| 40 | **n-Propanol** | 150 | 5.4 | 1 | 0.179 | 0.090 | 0 | 0.271 | 16 | 17.8 | [131] |
| 41 | **3-Bromopentane** | 167 | 8.4 | 1 | 0.183 | 0.091 | 0 | 0.277 | 18 | | |
| 42 | **2-methyl-1-propanol** | 172 | 6.32 | 1 | 0.188 | 0.094 | 0 | 0.283 | 19 | | |
| 43 | **Selenium** | 496 | 5 | 1 | 0.189 | 0.095 | 0 | 0.288 | 55 | 68.5 | [132] |
| 44 | **Butyronitrile** | 161 | 5.02 | 1 | 0.199 | 0.099 | 0 | 0.300 | 19 | 15.8 | [143] |
| 45 | **cis-/trans-Decalin** | 231 | 9.46 | 1 | 0.202 | 0.101 | 0 | 0.305 | 27 | | |
| 46 | **Ethanol** | 160 | 4.93 | 1 | 0.206 | 0.103 | 0 | 0.311 | 19 | 23 | [50] |
| 47 | **Methanol** | 172 | 3.85 | 1 | 0.209 | 0.105 | 0 | 0.314 | 21 | 36 | [50] |
| 48 | **Ethylene glycol** | 260 | 11.86 | 1 | 0.210 | 0.105 | 0 | 0.320 | 32 | 36 | [50] |
| 49 | **m-Xylene** | 225 | 11.56 | 1 | 0.221 | 0.111 | 0 | 0.311 | 29 | 28.5 | [105] |
| 50 | **ZnCl$_2$** | 590 | 6.7 | 1 | 0.180 | 0.090 | 0 | 0.253 | 62 | | |
| 51 | **B$_2$O$_3$** | 723 | 19.28 | 1 | 0.129 | 0.073 | 0 | 0.191 | 62 | 68.5 | [111] |

## 10. Stable-glass supercluster nucleation rates between T$_K$ and T$_g$

The Gibbs free energy change for a stable-glass nucleus formation is no longer given by (15) and is equal to (41) because the quantity $\Delta H_m/V_m*\theta$ is eliminated from the Gibbs free energy change leading to this new liquid state:

$$\Delta G_{2\lg}(R,\theta) = \frac{\Delta H_m}{V_m}(-\Delta\varepsilon_{\lg})4\pi\frac{R^3}{3} + 4\pi R^2 \frac{\Delta H_m}{V_m}(1+\Delta\varepsilon_{\lg})(\frac{12k_B V_m \ln K_{ls}}{432\pi \times \Delta S_m})^{1/3} \,. \quad (41)$$

The critical radius and the thermally-activated critical barrier are given by (42) and (43) instead of (9) and (10):

$$R^*_{2\lg} = \frac{-2(1+\Delta\varepsilon_{\lg})}{-\Delta\varepsilon_{\lg}} \left(\frac{V_m k_B \ln(K_{\lg})}{36\pi \Delta S_m}\right)^{1/3}, \quad (42)$$

$$\frac{\Delta G^*_{2\lg}}{k_B T} = \frac{12(1+\Delta\varepsilon_{\lg})^3 \ln(K_{\lg})}{81(-\Delta\varepsilon_{\lg})^2 (1+\theta)}. \quad (43)$$

The critical barrier in (43) is high because the coefficient $\Delta\varepsilon_{\lg}$ is always small in all liquids. Then, the stable glass phase cannot directly grow from the critical radius and is formed by homogeneous formation of numerous tiny superclusters percolating, interpenetrating, and then growing by reduction of their surface energy with the time increase. The Gibbs free energy change associated with the formation of a stable-glass nucleus of radius R containing n atoms is equal to (44) instead of (17):

$$\Delta G_{n\lg}(n,\theta,\Delta\varepsilon_{n\lg}) = \Delta H_m \frac{n}{N_A}(-\Delta\varepsilon_{n\lg}) + \frac{(4\pi)^{1/3}}{N_A}\Delta H_m(1+\Delta\varepsilon_{n\lg})\left[\frac{N_A k_B \ln(K_{\lg})}{36\pi \Delta S_m}\right]^{1/3}(3n)^{2/3}. \quad (44)$$

The energy saving coefficient $\Delta\varepsilon_{n\lg}$ of a glass nucleus of radius R containing n atoms is given by (45):

$$\Delta\varepsilon_{n\lg} = \Delta\varepsilon_{n\lg 0}\left[1 - 2(a-1) - \frac{9\theta^2}{2\theta_g^2}\left(\frac{1}{a} - \frac{2}{3}\right)\right], \quad (45)$$

where the coefficient $\Delta\varepsilon_{\lg 0} = -0.5 \times \theta_g$ in (32) has been replaced by $\Delta\varepsilon_{n\lg 0}$, which is proportional to the complementary Laplace pressure and to 1/R when the inequality n ≥ 147 is respected. The value of $\Delta\varepsilon_{n\lg 0}$ is predicted using (46) at $\theta = 0$:

$$\Delta\varepsilon_{n\lg 0} = \frac{\Delta\varepsilon_{\lg 0} \times R^*_{2\lg}}{R} = \Delta\varepsilon_{\lg 0}\left[\frac{n_c}{n}\right]^{1/3}, \quad (46)$$

where $\Delta\varepsilon_{\lg 0}$ is equal to $-0.5 \times \theta_g$ below $T_g$. At lower values of n, the energy saving is weakened by quantification and $\Delta\varepsilon_{n\lg 0}$ is strongly reduced [26,144]. In this particular case, the exact molar volume of superclusters being unknown, $\Delta\varepsilon_{n\lg 0}$ is better determined from the nucleation temperature of the stable-glass phase which occurs at the Kauzmann temperature $T_K$.

The values of $\ln K_{\lg}$ or $\ln K_{\lg}$ are calculated with (1) knowing that the Zeldovich factor $\Gamma$ is defined by (47), $n_c$ by (20) or (48) and $\Delta G^*/k_B T$ by (21) or (43), and that the transient times of nucleation at $T_g$ or at the nose temperature $T_n$ of the TTT diagram of crystallization above $T_g$ are close to 50 s:

$$\Gamma = \left(\frac{1}{3\pi n_c^2} \frac{\Delta G_2^*}{k_B T}\right)^{1/2}, \tag{47}$$

$$n_c = \frac{4\pi (R_{2lg}^*)^3}{3} \frac{N_A}{V_m} = \frac{8 N_A (1+\Delta\varepsilon_{lg})^3}{27} \frac{k_B \ln(K_{lg})}{\Delta S_m (\Delta\varepsilon_{lg})^3}. \tag{48}$$

The thermal variations of $\ln K_{ls}$ and $\ln K_{lg}$ in (3) respectively depend on $B_m/(T-T_{0m})$ and $B_g/(T-T_{0g})$, which are equal and deduced at $T_g$ from measurements of viscosity above and below $T_g$.

The glass and crystal steady-state nucleation times $t_{sn}$ given by (23) depending on the K value can be calculated as a function of the temperature when the effective thermally-activated energy barrier $\Delta G_{neff}/k_B T$ given by (22) or (49) is known:

$$\frac{\Delta G_{neff}}{k_B T} = \frac{12(1+\Delta\varepsilon_{nlg})^3 \ln(K_{lg})}{81(-\Delta\varepsilon_{nlg})^2 (1+\theta)} - \frac{\Delta G_{nlg}}{k_B T}, \tag{49}$$

where $\Delta G_{nlg}$ is given in (44). The crystallized or vitreous superclusters can grow beyond their own initial radius R when (23) or (50) is respected:

$$\ln(J_n v t_{sn}) = \ln(K_{lg} v t_{sn}) - \frac{\Delta G_{neff}}{k_B T}. \tag{50}$$

The n-atom supercluster formations occur in the sample when their nucleation time is evolved. The atom number n in spherical superclusters is chosen equal to the following stable magic numbers which are considered in an icosahedral structure for metals with face-centered cubic lattices: 13, 55, 147, 309, and 561 [47]. The nucleation rate logarithm of n-atom superclusters $\ln J_n = -\ln(v.t_{sn})$ is calculated without knowing the stable-glass domain volume v and the nucleation time $t_{sn}$ because the maxima of nucleation rates occur at the Kauzmann temperature $T_K$ and leads to a whole transformation of the liquid.

The homogeneous nucleation rates $\ln J_n$ of n-atom superclusters ready for growth in three BMG; $Pd_{43}Ni_{10}Cu_{27}P_{20}$ N°2, $Pt_{57.3}Cu_{14.6}Ni_{5.3}P_{22.8}$ N°7, $Cu_{47}Ti_{34}Zr_{11}Ni_8$ N°15, and four glasses; indomethacin G. N°8, $A_2Se_3$ G. N°17, diethylphthalate G. N°32 and selenium N°43 represented in Figure 7 are calculated using the parameters given in Table 6 and (50). The enthalpy saving is given by (32) and all results presented here are obtained without introducing a constant equilibrium enthalpy saving below $T_K$, in order to show that the model directly leads to the value of $T_K$. The same maximum at $T_K$ is still observed when a constant enthalpy change is introduced below $T_K$. A nucleation rate equal to exp (20.7)/m$^3$/s would transform a liquid volume of 1 mm$^3$ into a stable glass in an additional time of 1 s if the glass domain could attain this volume by nucleus growth beyond the critical radius. It is not possible because the critical energy barrier given in (43) is always too high. Elementary clusters containing 13, 55, 147, 309, 561 and 923 atoms were studied. The numbers n = 147 or 309 lead directly to the highest maxima of $\ln J_n$ at $T_K$ using $\Delta\varepsilon_{nlg0}$ values obeying (46). The model applied to four glasses works without using any adjustable parameter for n = 147 and 309. The atom numbers n have been chosen in Figure 7 as the number n inducing the largest nucleation rate at $T_K$. The steady-state nucleation rate depends on the fusion enthalpy $\Delta H_m$ as shown by (44) and (17). The examples given in Figure 7 cover a broad distribution of fusion heats and, consequently, various nucleation rates in glass-forming melts. There is no stable glass nucleus being formed at $T_g$. The supercluster nucleation rate has a maximum at $T_K$ in all these examples. The transition at $T_g$ cannot be described by supercluster nucleation having a surface energy in spite of the knowledge of the enthalpy difference $\Delta\varepsilon_{lg}\times\Delta H_m/V_m$ between undercooled liquid and stable-glass phase and the specific heat jump prediction. The transition at $T_g$ is a liquid-liquid transition characterized by the enthalpy change that is predicted in (32). Other liquid-liquid transition models involving superclusters of liquid nature are more successful to describe them at the microscopic scale [6-8,10-15].

**Table 6. Parameters used to calculate the nucleation rates $\ln J_n$, the nucleation times t and $t_{sn}$ of stable vitreous phases.** The units are based on meter, kelvin, joule and second. The entropy $\Delta S_m$ is given per g.atom and $V_m$ in $m^3$ per mole. These droplets give rise to very tiny stable-glass domains and to maxima of nucleation rates at $T_K$, as shown in Figure 7. The critical number of cluster atoms $n_c$ at the melting temperature $T_m$, the nucleation rate logarithm $\ln(J_n/m^3/s)$ at $T_K$ of n-atom elementary superclusters, the extrapolated negative critical energy saving coefficient $\Delta\varepsilon_{lg0}$ at $T_m$ as shown in Figure 1, and the energy saving coefficient $\Delta\varepsilon_{nlg0}$ at $T_m$ ($\theta = 0$) of n-atom superclusters are given. The $\ln K_{lg}$ value is determined assuming that the transient relaxation time at $T_g$ is equal to 50 s.

| Glass | $\ln K_{lg}$ $T=T_g$ | $B/(T_g-T_{0g})$ | $-\Delta\varepsilon_{lg0}$ | $T_{0g}$ | $-\Delta\varepsilon_{nlg0}$ | $\Delta S_m$ | $V_m$ $\times 10^6$ | n | $n_c$ | $T_K$ | $T_g$ | $\ln J_n$ |
|---|---|---|---|---|---|---|---|---|---|---|---|---|
| $Pd_{40}Ni_{10}Cu_{30}P_{20}$ | 64.3 | 35.4 | 0.141 | 365 | 0.465 | 8.74 | 8 | 147 | 5274 | 510 | 576 | 47.1 |
| $Pt_{57.3}Cu_{14.6}Ni_{5.3}P_{22.8}$ | 63.1 | 31.9 | 0.19 | 327 | 0.5 | 13.9 | 10.1 | 55 | 1176 | 419 | 509 | 29.7 |
| $Cu_{47}Ti_{34}Zr_{11}Ni_8$ | 63 | 34.7 | 0.201 | 437 | 0.4 | 10.02 | 10.2 | 55 | 1448 | 541 | 673 | 15.6 |
| Indomethacin | 65 | 38 | 0.132 | 203 | 1.029 | 2.22 | 271 | 147 | 26439 | 285 | 318 | 44.5 |
| $As_2Se_3$ | 64.7 | 36 | 0.211 | 296 | 0.406 | 5.15 | 81.4 | 309 | 2089 | 431 | 462 | 47.8 |
| Diethylphthalate | 64.7 | 36 | 0.167 | 113 | 0.716 | 2.25 | 198 | 309 | 11608 | 152 | 178 | 36.9 |
| Selenium | 62.6 | 36 | 0.189 | 198 | 0.52 | 10.08 | 16.5 | 55 | 1643 | 254 | 309 | 25 |

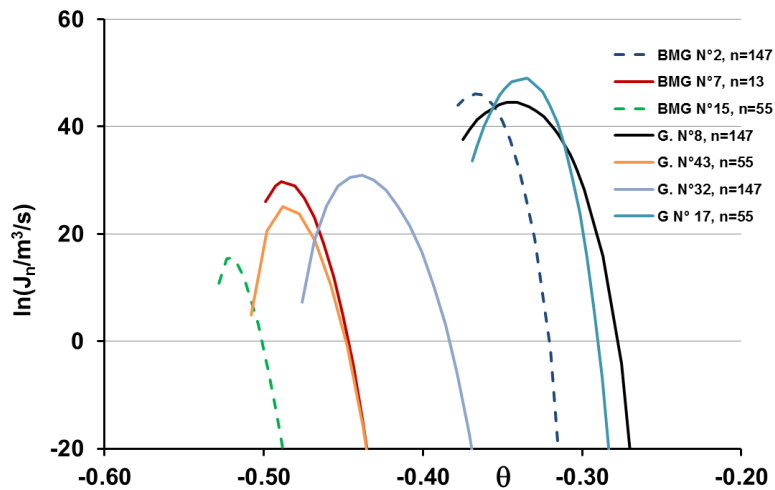

**Figure 7. The n-atom supercluster nucleation rate logarithms $\ln J_n$ of seven glass-forming melts versus $\theta = (T-T_m)/T_m$.** They are plotted versus the reduced temperature below $\theta_g$. The unit of $J_n$ is $m^{-3}.s^{-1}$. The nucleation rates of stable-glass superclusters are negligible at $T_g$, while they are high and maximum at $T_K$. A liquid-to-liquid transition occurs at $T_g$ without stable-glass supercluster formation in $Pd_{43}Ni_{10}Cu_{27}P_{20}$ BMG N°2, $Cu_{47}Ti_{34}Zr_{11}Ni_8$ BMG N°15, indomethacin G. N°8, selenium G. N°43, diethylphthalate G. N°32, $As_2Se_3$ G. N°17, and $Pt_{57.3}Cu_{14.6}Ni_{5.3}P_{22.8}$ BMG N°7.

## 11. Fragile-to-fragile liquid transition at $T_g$ always occurring above $T_m/2$

All the fragile glass-forming melts in Tables 2 and 3 have a transition temperature $T_g$ larger than $T_m/2$. A strong-to-fragile liquid transition only exists when $\varepsilon_{lgs0}$ is smaller than 1.25, as shown in Table 1.1. In all the given examples, the ideal glass transition temperature $T_{0g}$ is always lower than $T_K$. The Kauzmann temperature cannot be lower than $T_{0g}$. A transition at $T_g = T_m/2$ would lead to $T_{0g} = T_K = 0.3545 \times T_m$, $T_{0m} = 0.423\, T_m$ and limiting values equal to 1.25 for $\varepsilon_{lg0}$ and 1.5 for $\varepsilon_{lso}$. This property explains why some fragile glass-forming liquids do not undergo a visible liquid-to-liquid transition before being crystallized by heating them at temperatures a little higher than $T_m/2$ [145,146]. When a visible transition temperature $T_g$ is lower than $T_m/2$, as shown in Table 1-1 for $Au_{77}Ge_{13.6}Si_{9.4}$, a fragile-to-strong liquid transition exists for $T_m/3 < T_{0m} < 0.3715 \times T_m$. There is no liquid-liquid transition in any undercooled melt for $0.3715 \times T_m < T_{0m} < 0.423 \times T_m$ and an amorphous state is observed below the crystallization temperature.

## 12. Hidden freezing at $T_g$ before relaxation of quenched liquids

The specific heat of a quenched liquid is always assumed as being continuous below $T_g$ before relaxation. The Kauzmann temperature is extrapolated using the specific heat thermal variation measured from $T_g$ to $T_m$. This extrapolation is in contradiction with the linear decrease of the relaxed enthalpy with temperature which reveals that the specific heat is nearly constant below $T_g$. In fact, there is a slope change of the specific heat at $T_g$. This change is very often small for samples obeying the scaling law above $T_g$ and often large in Table 5 for samples having a latent heat $L^-$ delivered at $T_g$ by cooling. A signature of a freezing transition at $T_g$ exists without being fully accomplished before enthalpy relaxation, as shown in Figures 8, 9 and 10. The $Pd_{43}Cu_{27}Ni_{10}P_{20}$ undercooled liquid which follows the scaling law above $T_g$ has a specific heat slope decreasing at $T_g$, whereas that of $Zr_{44}Ti_{11}Ni_{10}Cu_{10}Be_{25}$ and maltitol are increasing.

The theoretical and experimental Kauzmann temperatures of maltitol are not the same as shown in Figure 10 [147]. The calculated one is weakened by a reduction of the available entropy below $T_g$ due to the latent heat $L^-$ delivered during cooling at $T_g$. Many discrepancies between calculated and extrapolated values of $T_K$ are explained by a reduction of the available entropy due to the existence of a latent heat $L^-$ at $T_g$.

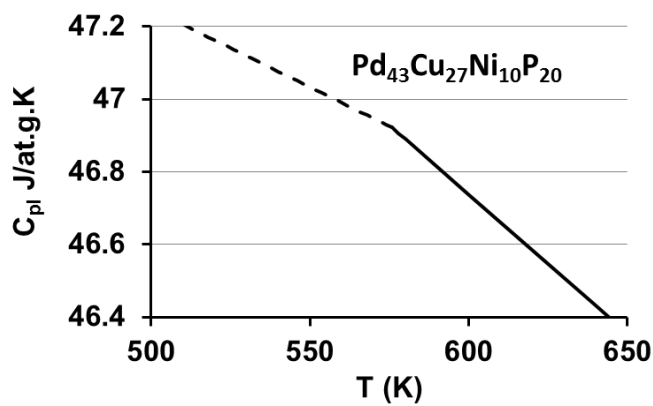

**Figure 8. Specific heat of the undercooled liquid $Pd_{43}Cu_{27}Ni_{10}P_{20}$.** The undercooled liquid specific heat $C_{pl}$ of BMG N°2 is plotted versus temperature above $T_g = 576$ K using experimental results [79]. The specific heat

between $T_g$ and $T_K$ is calculated by adding the specific heat change given by (34) to the experimental values $C_{pg}$ of the glass phase. There is a weak change of the slope at $T_g$ without exothermic latent heat. This explains why the calculated and extrapolated values of $T_K$ are about the same.

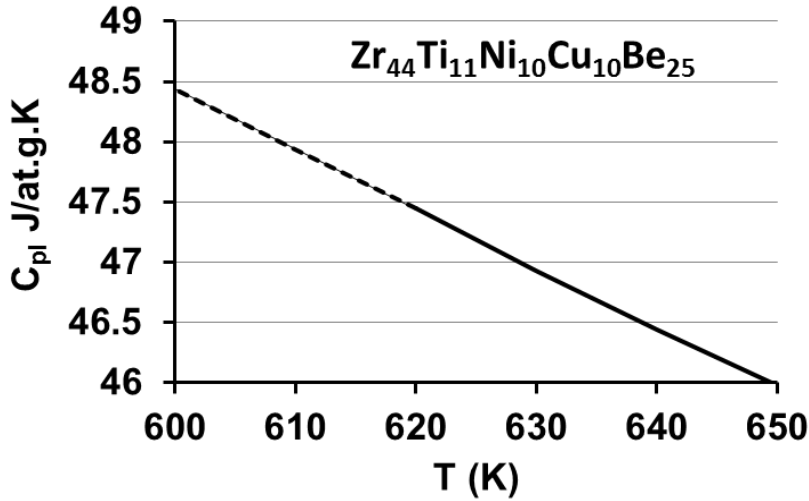

**Figure 9. Specific heat of the undercooled liquid $Zr_{44}Ti_{11}Ni_{10}Cu_{10}Be_{25}$.** The undercooled liquid specific heat of BMG N°3 is plotted versus temperature above $T_g$ using known experimental results [76]. Below $T_g$ down to $T_K$, the specific heat of the new liquid state has been calculated adding (34) and 0.36 J/at.g.K to the crystallized phase specific heat instead of introducing a complementary slope corresponding to the difference of specific heat between the glass and crystallized states ($C_{pg}$-$C_{px}$). The calculated and experimental values of $\Delta T_K$ are equal.

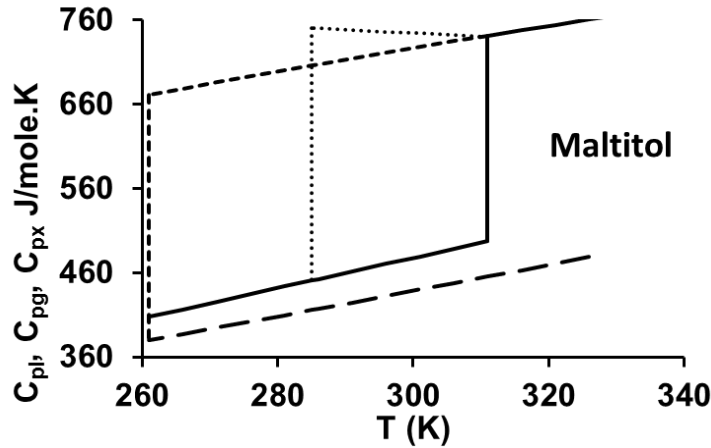

**Figure 10. The undercooled liquid specific heat of maltitol.** The undercooled liquid specific heat of Glass N°5 is plotted versus temperature above and below $T_g$ using known experimental results (continuous line) [146]. An extrapolation of $\Delta C_{pl}$ below $T_g$ leads to a Kauzmann temperature of 261 K. The specific heat below $T_g$ is calculated by adding (34) to the measured glass specific heat (point line). There is a slope increase accompanying the freezing transition before enthalpy relaxation. The Kauzmann temperature deduced from (32) is smaller and equal to 285 K.

## 13. Transformation of the stable-glass indomethacin in undercooled liquid at $T_g$

The specific heat of four indomethacin samples has been measured and compared [148,149,150]. The first one is a stable glass which has been obtained by physical vapor deposition. The second one is an ordinary glass and the two others have been isothermally aged below $T_g$ during 7 months and 37 days. The enthalpy of the aged and stable glass samples below $T_g$ are smaller than that of an ordinary glass [102,123]. The authors have claimed that their ageing times have not transformed indomethacin into stable glass. The enthalpy difference between an ordinary glass and the stable glass is equal to $4000 \pm 400$ J/mole and the predicted latent heat $L^+$ in Table 5 N°8 is smaller and equal to $0.066 \times \Delta H_m =$ 2600 J/mole. The specific heat difference between ordinary glasses and those submitted to ageing increases slightly with ageing time. The specific heat of a stable glass is also a little less than that of an ordinary glass. These observations show that the pseudo-equilibrium obtained when $\tau^{ns}$ is evolved after structural relaxation is not fully attained as expected from theoretical considerations, which have shown that the steady-state and transient nucleation times are mixed and not simply added [37]. The model used here predicts the specific heat difference between a quenched undercooled liquid and its stable glass phase instead of the difference between a quenched undercooled liquid and the new liquid phase in a pseudo-equilibrium state. The small experimental increase of $\Delta C_{plg}$, which is equal to $19 \pm 10$ J/mole.K, increases the enthalpy difference, and a corrected value of $19 \times \Delta T_K = 627$ J/mole.K reduces the observed latent heat at $T_g$ from 4000 to $3373 \pm 400$ kJ/mole.K [148]. The transition temperature $T_g$ observed after ageing is 7 K higher, as shown by specific heat jumps which are as large as that of the stable-glass phase [150](Fig. 6). The enthalpy recovery at $T_g$ which has been previously relaxed by the undercooled melt at room temperature during ageing is equal to about $\Delta C_{plg} \times 20$ K. This enthalpy has to be reinjected in the sample at the thermodynamic glass transition. The endothermic latent heat of this aged sample is equal to 2900 J/mole, in agreement with the observed enthalpy excess. The thermodynamic transition temperature is close to 325 K for both aged, relaxed samples and stable glass samples instead of 318 K as measured for ordinary glasses. It is so because the nucleation time of the liquid phase in the glass phase seems to be minimum at this temperature and equal to about 4000 s in agreement with nanoscale specific heat measurements [148]. This relaxation time is larger than 50 s because the stable-glass density and consequently the energy barrier for atom diffusion from the vitreous state to the undercooled melt is larger [39].

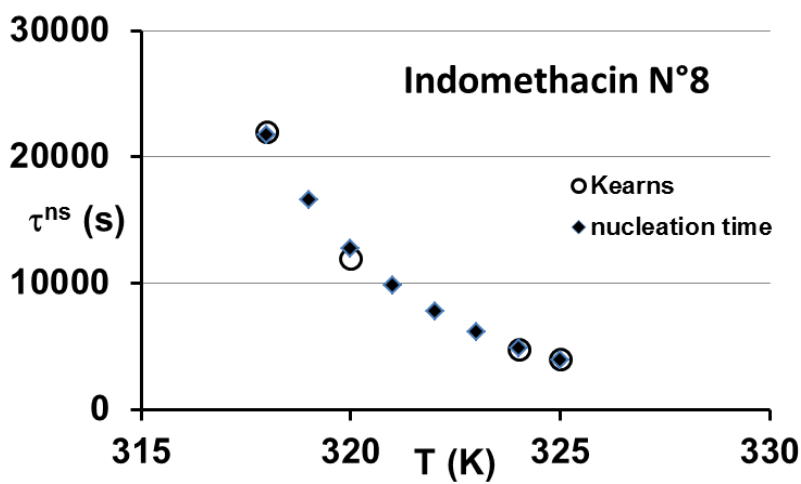

**Figure 11. The transient nucleation time-lag $\tau^{ns}$ of undercooled liquid droplets containing the critical atom number in the stable glass phase**. This calculated transient nucleation time-lag $\tau^{ns}$ is plotted versus

32temperature. Experimental points noted Kearns are found in [148] (Fig. 6). The samples are thin films of stable glass phase which have a thickness of 2900 nm, a volume v = 1.13 mm$^3$ and are submitted to annealing at temperatures lower than $T_g$ and equal to 318, 320, 324 and 325 K.

The new thermodynamic parameters are: $\Delta\varepsilon_{lg0}$ = 0.124, $\varepsilon_{ls0}$ = 1.752, $\varepsilon_{lg0}$ = 1.628, $\theta_{0g}$ = −0.518, $\theta_{0m}$ = −0.439. The Gibbs free energy change has reversed its sign and is given in (29) with $(\varepsilon_{ls0}-\varepsilon_{lg0}) = \Delta\varepsilon_{lg0}$. All stable-glass superclusters have a negligible steady-state nucleation time at $T_g$. The transient nucleation time-lag $\tau^{ns}$ in (1) is chosen as being equal to 3910 s at $T_g$ = 325 K corresponding to $\ln K_{gl}$ = 64.57 in (3), $B/(T_g-T_{0m})$= 38.33 in (3), the Zeldovitch factor $\Gamma$ = 5.5×10$^{-4}$, $\Delta G^*_{2lg}/k_BT$ = 2754 and $V_m$ = 271.05×10$^{-6}$ m$^3$ for 41 atoms per indomethacin molecule. A good agreement with experimental values of $\tau^{ns}$ is obtained as shown in Figure 11 [147](figure 6). The time dependence of the melting temperature of the stable glass is due to the thermal variation of the transient nucleation time-lag $\tau^{ns}$.

## 14. Conclusions

A new model introduces an enthalpy saving at $T_g$ describing the equilibrium property changes of many fragile glass-forming melts below $T_g$ only knowing $T_g$, $T_m$ the melting temperature, and $\Delta H_m$ the melting heat. The specific heat jump at $T_g$, the reduction from $T_{0m}$ to $T_{0g}$ of the temperatures at which the enthalpy savings associated with crystal nucleus formation would be extrapolated to zero from above and below $T_g$, the Kauzmann temperature $T_K$, the enthalpy saving between $T_g$ and $T_K$, the relaxed enthalpy, its ultimate value at $T_K$, the latent heats associated with liquid-stable-glass transition and TTT diagrams of vitreous and crystallized phase nucleation are predicted. The transition at $T_g$ is, in a first step, a liquid-liquid transition with a pseudo-equilibrium time which strongly increases when the temperature decreases below $T_g$.

The time dependence of $T_g$ measured varying cooling and heating rates is due to this incubation time which is viewed as a transient nucleation time $\tau^{ns}$ in undercooled liquids preparing the phase transformation. The incubation time at $T_g$ is equal to about 50 s and is the same as that observed in TTT diagrams at the crystallization nose temperature above $T_g$ in spite of a large change of the viscosity between these two temperatures. It is proportional to the reverse of the constant K defining the transient nucleation time $\tau^{ns}$ in (1). A mean value of $\ln K$ = 63.9 ± 1.3 is obtained at $T_g$ for all liquids listed in Table 6.

Fragile-to-fragile and strong-to-fragile liquid transitions are observed. Strong liquids are transformed into stronger liquids and stable-glasses by cooling below $T_g$ without latent heat. In all liquids, an enthalpy is always relaxed below $T_g$ during the time-lag of transient nucleation. The small specific heat jumps at $T_g$ in strong liquids are used to determine their temperature $T_{0m}$ lower than $T_m/3$, assuming that the new value of $T_{0g}$ becomes equal to zero at 0 K. The model also predicts the absence of fragile-to-fragile liquid transition near $T_m/2$ when $T_{0m}$ is larger than 0.3715×$T_m$ and smaller than 0.423×$T_m$.

The energies savings $\varepsilon_{ls}(\theta)\times\Delta H_m/V_m$ and $\varepsilon_{lgs}(\theta)\times\Delta H_m/V_m$ are enthalpy excesses per unit volume of undercooled liquids above and below $T_g$ as compared to that of the crystallized state while $\Delta\varepsilon_{lg}(\theta)\times\Delta H_m/V_m$ is the enthalpy excess of the undercooled liquid as compared to that of the stable-glass phase with $\Delta\varepsilon_{lg}(\theta)$ being equal to $[\varepsilon_{ls}(\theta) - \varepsilon_{lgs}(\theta)]$. These quantities induce different Laplace pressures on the superclusters which may lead to the condensation of a new phase from the undercooled melt. The energy saving $\varepsilon_{ls}(\theta) \times \Delta H_m$ giving rise to crystallization above $T_g$ is transformed below $T_g$ in $\varepsilon_{lgs}(\theta) \times \Delta H_m$ respecting $\varepsilon_{lgs} < \varepsilon_{ls}$ with a minimized value of $\varepsilon_{lgs}$. The new liquid phase below $T_g$ cannot be transformed in crystals because the nucleation time of a critical supercluster is much too long. The stable glass phase formation is driven by an enthalpy change at $T_g$ equal to $\Delta\varepsilon_{lg}(\theta) \times \Delta H_m / V_m$. The specific heat difference between an undercooled melt and the stable-glass phase is predicted between $T_K$ and $T_g$. The supercluster homogeneous nucleation temperatures $T_1$ above $T_g$ and $T_2 = T_g$ occur far below the crystallization temperature when the energy saving coefficients $\varepsilon_{ls0}$ and $\varepsilon_{lg0}$ at $T_m$ are minimum, $\varepsilon_{lg0}$ being always equal to $1.5 \times \theta_g + 2$ and $\varepsilon_{ls0}$ to $\theta_g + 2$ in about 70% of all fragile glass-forming melts where $\theta_g = (T_g - T_m)/T_m$. Their temperatures $T_{0g}$ and $T_{0m}$ follow scaling laws depending on $\varepsilon_{lg0}$ and $\varepsilon_{ls0}$, their specific heat jump at $T_g$ being equal to $1.5 \times \Delta H_m / T_m$, and their reduced Kauzmann temperature square respecting $\theta^2_K = 5 \times \theta^2_g / 3$.

The critical stable-glass nucleus is submitted to a Laplace pressure change $\delta p$ accompanying the enthalpy change $-V_m \times \delta p = -\Delta\varepsilon_{lg} \times \Delta H_m$ per mole below $T_g$. The critical energy saving coefficients $\varepsilon_{ls}$, $\varepsilon_{lgs}$ and $\Delta\varepsilon_{lg}$ are linear functions of $\theta^2$ instead of $\theta = (T-T_m)/T_m$. The energy saving coefficients $\varepsilon_{nlgs0}$ and $\varepsilon_{nls0}$ at $T_m$ of a n-atom supercluster are proportional to its reverse radius when $n \geq 147$ and is quantified for $n < 147$. Any n-atom supercluster formed below $T_m$ is not submitted to premelting because it has an energy saving coefficient depending on $\theta^2$ and consequently its surface atoms have the same melting temperature $T_m$ than the core ones.

The critical energy barriers of melts are too high below $T_g$ for having a single supercluster giving rise by growth to an infinite supercluster at any temperature smaller than $T_g$. The equilibrium glass phase is obtained by homogeneous formation, percolation and interpenetration of elementary supercluster multitude containing magic atom numbers filling all the space during very long nucleation times. The critical energy barrier associated with these elementary superclusters depends on their atom number n and is reduced by the large Gibbs free energy change associated with their formation. Their nucleation times and TTT diagrams are predicted between $T_K$ and $T_g$ in appendix A. The stable vitreous phase is mainly formed at the Kauzmann temperature $T_K$ because the nucleation rate of superclusters which are ready to grow has a pronounced maximum at this temperature where the nucleation times are minimum values. The liquid-to-liquid transition at $T_g$ is not induced by stable-glass supercluster formation and is a direct consequence of the enthalpy change.

The thermodynamic transitions at $T_g$ cannot be described as second-order or first-order transitions. All fragile and strong liquids produce relaxed enthalpy after quenching followed by annealing temperatures between $T_K$ and $T_g$. An endothermic latent heat is recovered in fragile liquids at $T_g$ during heating which depends on the annealing temperature below $T_g$ and is only equal to the relaxed enthalpy when a stable glass phase has not been formed. The stable glass phase formation also produces an exothermic enthalpy at $T_K$ which is recovered at $T_g$ by a complementary endothermic latent heat. In strong liquids, there is no latent heat accompanying the stable-glass formation and the transition at $T_g$ is a liquid-liquid transition after a minimum of relaxation time below $T_g$ in agreement with descriptions of the formation of dynamical fractal structures near a percolation threshold. In many glass-forming melts, there is no first-order transition because there is no exothermic latent heat produced at $T_g$ during cooling in spite of the presence of an endothermic latent heat during heating. A second-order phase transition character exists because the specific heat difference between liquid and glass is successfully calculated in all liquids using the first-derivative of the enthalpy saving. The glass-forming melt is sometimes so fragile that a reversible latent heat exists at $T_g$ which is only a fraction of the available enthalpy. In these liquids, the total endothermic latent heat at $T_g$ is much larger than the exothermic latent heat obtained by cooling when it exists.

**Appendix A**
 **TTT diagrams of several liquids in stable-glass phase between $T_K$ and $T_g$**

The parameters used to calculate the nucleation times are given in Table 6 for three BMG: $Pd_{43}Ni_{10}Cu_{27}P_{20}$ N°2, $Pt_{57.3}Cu_{14.6}Ni_{5.3}P_{22.8}$ N°7 and $Cu_{47}Ti_{34}Zr_{11}Ni_8$ N°15 and four glasses: indomethacin N°8, $A_2Se_3$ N°17, diethylphthalate N°32 and selenium N°43. The nucleation times t and $t_{sn}$ of these seven glasses are plotted as a function of temperature in Figures A1, A2, A3, A4, A5, A6, and A7 respectively. The total time $t = \tau^{ns} + \pi^2/6 \times t_{sn}$ includes the transient relaxation time $\tau^{ns}$ starting from the quenched liquid state, while $t_{sn}$ is the steady-state nucleation time of the stable-glass domain of volume v having a minimum at $T_K$. The critical energy barrier being much too high, the volume v is chosen equal to the volume of the n-atom cluster instead of the sample volume in order to have the minimum of t very close to $T_K$. This assumption is in agreement with the description of the glass-state by molecular dynamics simulations "as composed of tiny icosahedral-like clusters, most of which touching or interpenetrating yielding a microstructure of polyicosahedral clusters that follow a specific sequence of magic numbers"[23]. The Zeldovitch factor is calculated with (47) using $\Delta G^*_{nm}/k_BT$ instead of $\Delta G^*_{2lg}/k_BT$ and $n_c$ equal to the number n. The main uncertainty on the nucleation time t comes from the uncertainty on the value of $B/(T_g-T_{0g})$ chosen at $T_g$, given in Table 6 and from the transient nucleation time-lag $\tau^{ns}$ of stable-glass superclusters which could be much larger than 50 s as shown for indomethacin in part 12. It explains why the minimum of t sometimes occurs at a temperature a little larger than $T_K$. It is shown here that the nucleation rates of these elementary clusters are sufficiently large after ageing to produce this type of microstructure knowing that the growth around these nuclei could be strongly accelerated when the surface energy declines with touching and interpenetrating superclusters.

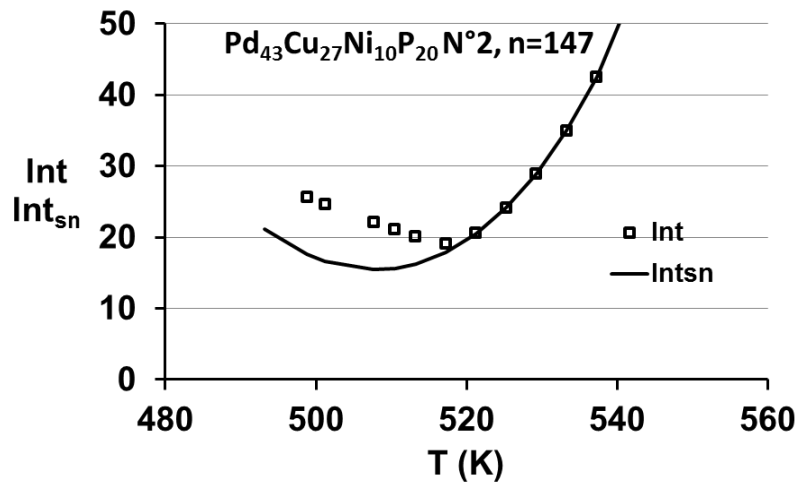

**Figure A1. TTT diagrams of $Pd_{43}Ni_{10}Cu_{27}P_{20}$.** The TTT diagrams of BMG N°2 represented by the logarithms of the total nucleation time t (s) and the steady-state nucleation time $t_{sn}$(s) are plotted versus temperature. The minima of $\ln t_{sn}$ and $\ln t$ occur at 510 K and 521 K respectively. The Kauzmann temperature is equal to 512 K. The supercluster volume logarithm $\ln(v/m^3)$ is equal to $-61.5$. It contains 147 atoms.

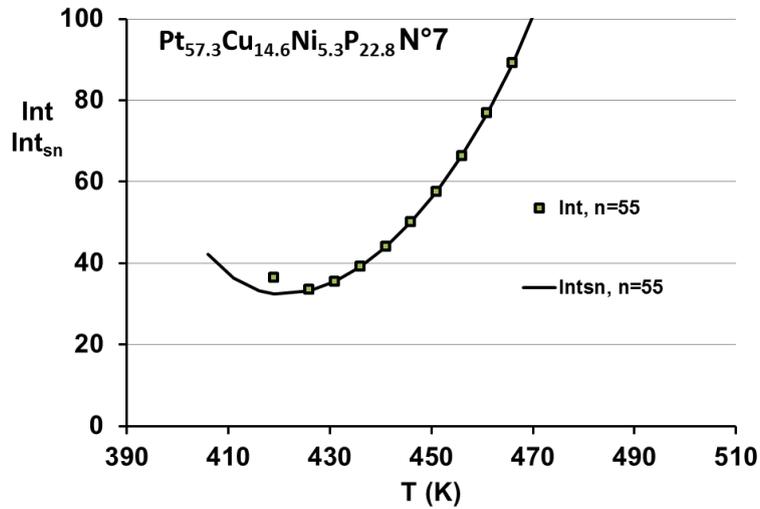

**Figure A2. TTT diagrams of $Pt_{57.3}Cu_{14.6}Ni_{5.3}P_{22.8}$.** The logarithms of the steady-state nucleation time $t_{sn}(s)$ and of the total nucleation time $t(s)$ of BMG N°7 are plotted versus temperature. The elementary cluster contains 55 atoms. The minimum of $t_{sn}(s)$ occurs at $T_K = 419$ K. The minimum of $t(s)$ occurs at $T = 426$ K. The supercluster volume logarithm is $\ln(v/m^3) = -62.2$.

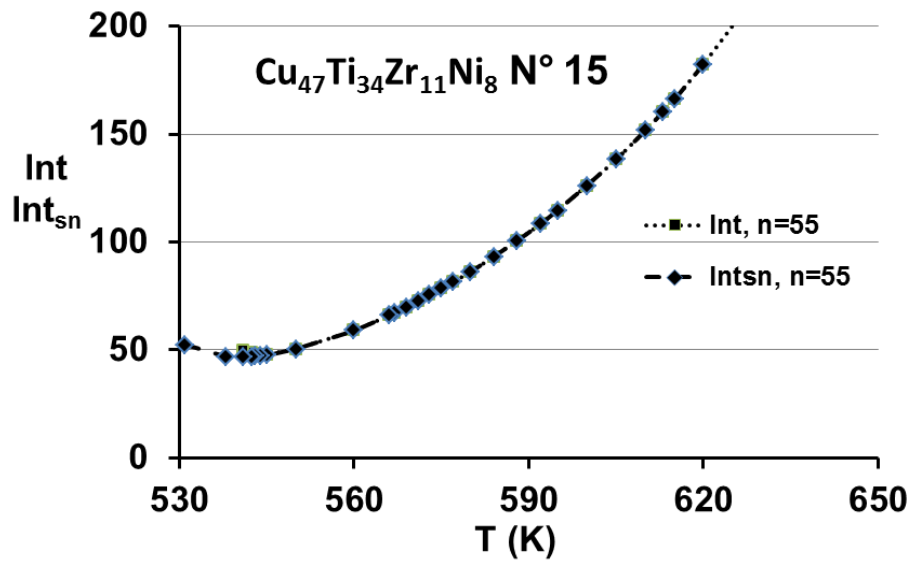

**Figure A3. TTT diagrams of $Cu_{47}Ti_{34}Zr_{11}Ni_8$ versus T**. The logarithms of the steady-state nucleation time $t_{sn}$ and the total nucleation time $t$ of BMG N°15 are plotted versus temperature. The elementary cluster contains 55 atoms. The minimum of $t_{sn}$ occurs at $T_K = 541$ K while that of $t$ occurs at 545 K. The supercluster volume logarithm is $\ln(v/m^3) = -62.3$.

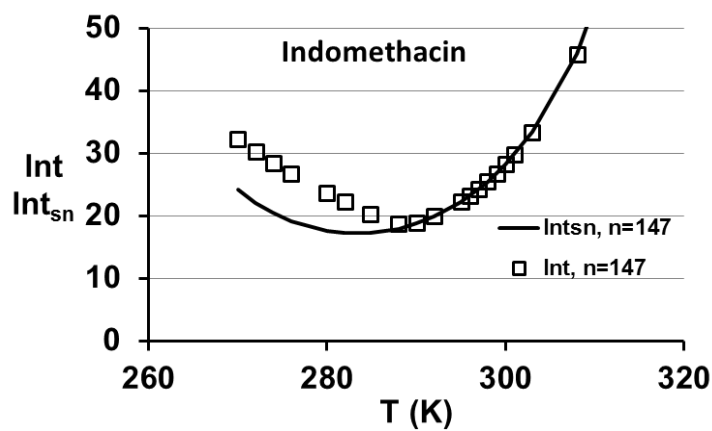

**Figure A4. TTT diagrams of Indomethacin**. The logarithms of the steady-state nucleation time $t_{sn}$ and the total nucleation time t of Glass N°8 are plotted versus temperature. The elementary supercluster contains 147 atoms. The minimum of $t_{ns}(s)$ occurs at T = 282 K while that of t(s) occurs at T = 288 K. The Kauzmann temperature is equal to 285 K. The supercluster volume logarithm is $\ln(v/m^3) = -61.8$. The indomethacin molecule contains 41 atoms.

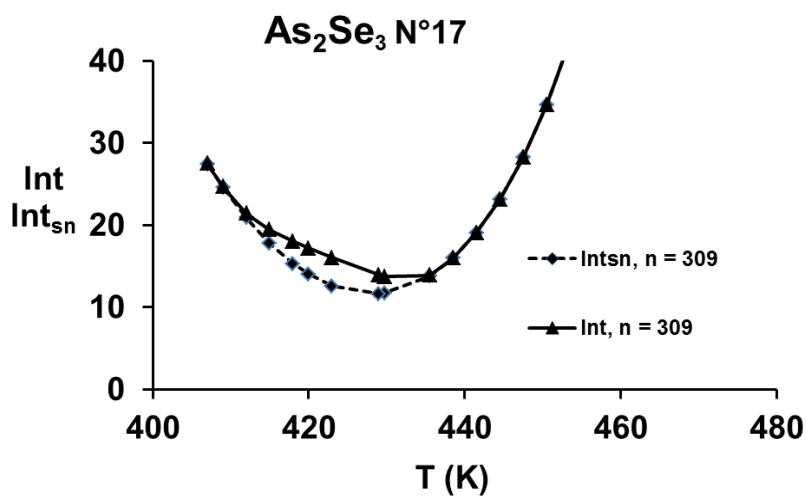

**Figure A5. TTT diagrams of $As_2Se_3$**. The logarithms of the steady-state nucleation time $t_{sn}$ and the total nucleation time t of Glass N°17 are plotted versus temperature. The elementary clusters contain 309 atoms. The minimum of $t_{sn}(s)$ and t(s) occurs at T = 430 K for n = 309. The supercluster volume logarithm is $\ln(v/m^3) = -59.5$. The calculated Kauzmann temperature is $T_K = 434$ K.

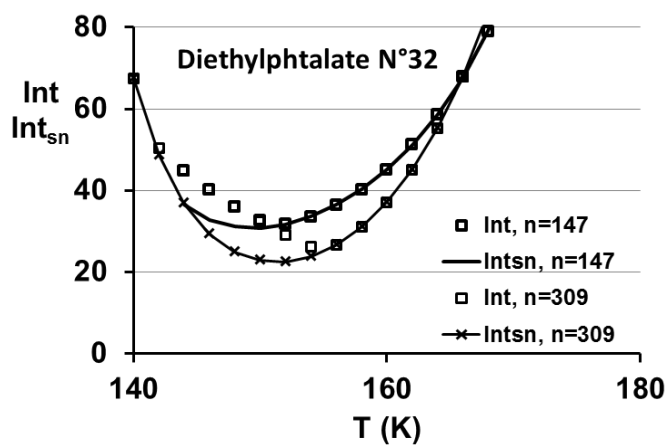

**Figure A6. TTT diagrams of diethylphthalate**. The logarithms of the steady-state nucleation time $t_{sn}$ and of the total nucleation time t of Glass N°32 are plotted versus temperature. The supercluster contains 147 and 309 atoms. The minima of $t_{sn}(s)$ and t(s) for n=309 occur at 152 K and $T_K$ = 154 K respectively. The supercluster volume logarithm is $\ln(v/m^3) = -61.8$. The diethylphthalate molecule contains 30 atoms.

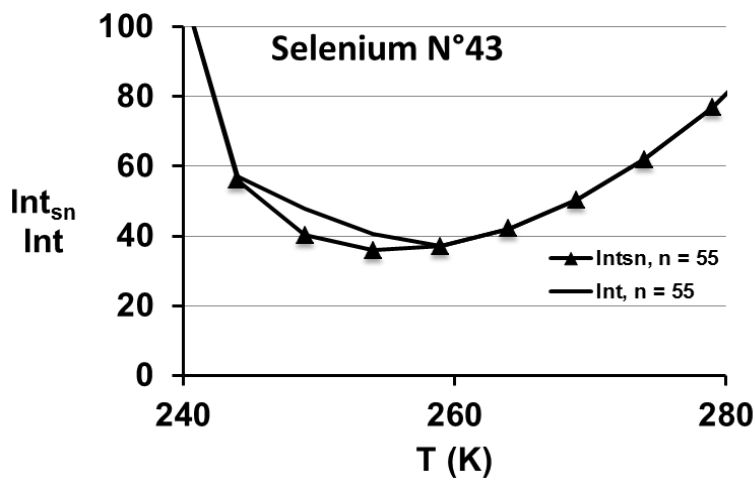

**Figure A7. TTT diagrams of Se**. The logarithms of the steady-state nucleation time $t_{sn}$ and the total nucleation time t of Glass N°43 are plotted versus temperature. The cluster contains 55 atoms. The minima of $t_{sn}(s)$ and t (s) occur at $T_K$ = 254 K and T= 259 K respectively. The supercluster volume logarithm is $\ln(v/m3) = -60.8$.

The nucleation rate of elementary clusters becomes larger at temperatures much smaller than $T_g$ and attains a maximum value at $T_K$, as shown in Figure 7. The formation of a stable glass phase by ageing is possible when the long formation time of an elementary supercluster is evolved and when (50) is respected authorizing its growth. All the liquids studied in Figure 7 could be transformed between $T_K$ and $T_g$ in times equal to the sum of structural relaxation time $\tau^{ns}$ and the formation time of the elementary supercluster. Studies are necessary to determine whether the transformation times by ageing at $T_K$ can become sufficiently small in some glasses. Some liquids such as $As_2Se_3$, $Pd_{43}Cu_{27}Ni_{10}P_{20}$ and Indomethacin having the largest nucleation rates are good candidates for such studies. The nucleation times of $As_2Se_3$ are the smallest and an ageing time of about 35 hours (within an uncertainty of about 35×7 hours) at T = 430 K could be sufficient to induce the stable-glass phase. This rough estimation is in agreement with the complementary latent heat in agreement predicted and measured after an ageing of 166 hours at 418 K [51]. The substrate temperature used for indomethacin physical vapor deposition has been varied from 265 to 305 K [44](Fig. 5). There is no more stable glass formation above 297 K in agreement with the rapid fall of the nucleation rate above $T_K$, as shown in Figure 7.

The stable glass phase cannot be obtained at $T_g$ by cooling because the nucleation rate logarithms are negative in all liquids. A liquid-to-liquid transition is nevertheless easily observed at $T_g$ because the time $\tau^{ns}$ necessary to attain the pseudo-equilibrium of the new liquid phase is of the order of 50 s. The enthalpy recovery of a quenched melt increases during cooling down to $T_K$ and is relaxed at each annealing temperature between $T_K$ and $T_g$ when the undercooled melt attains its pseudo-equilibrium, as shown in Figs 4-6. The whole undercooled liquid is transformed into a glass phase at $T_K$. The maxima of the elementary supercluster nucleation rate at $T_K$ are obtained without using any adjustable parameter for n ≥ 147 with $\Delta\varepsilon_{nlg0}$ being calculated with (46). The coefficient $\Delta\varepsilon_{nlg0}$ for n = 55 is adjusted by fixing the maximum nucleation rate at $T_K$. An exothermic latent heat has to be produced to undergo the transition at $T_K$ and the glass survives in this new equilibrium state from $T_K$ to $T_g$ marked by hysteresis cycles.

The glass phase is melted at $T_g$ with the help of the endothermic heat $L^+$. The enthalpies of $Pd_{43}Ni_{10}Cu_{27}P_{20}$, $As_2Se_3$ and indomethacin undercooled liquids at equilibrium are represented below $T_g$ in Figs 4-6. Along the recovered enthalpy line obtained by relaxation below $T_g$, the viscosity attains its pseudo-equilibrium value after an increase by a factor of 2 to 3 from its value in the quenched-liquid state before relaxation [64,76,78,93]. This viscosity relaxation shows that the undercooled liquid has already undergone a change into a frozen liquid state before relaxation. A time dependence of $T_g$ is observed during heating of rapidly quenched melts which have not had the time to attain the pseudo-equilibrium during cooling through $T_g$ because a minimum time of about 50 s is needed to undergo the transition [52,76,144].

The calculations of the enthalpy saving coefficients $\Delta\varepsilon_{lg}$ are only based on the knowledge of the glass transition $T_g$, the melting temperature $T_m$ and thermodynamic considerations related to scaling laws. The fusion enthalpy $\Delta H_m$ has to be known to calculate the latent heats and the ultimate enthalpies of the stable-glass phase.

## Appendix B
**Superclusters of 13 atoms governing the first-crystallization time of metallic glass-forming melts**

Superclusters submitted to various Laplace pressure give rise to stable- glass phase and others having different energy saving coefficients are acting as growth crystal nuclei far above $T_g$.

**Table B1. Parameters used to calculate the Time-Temperature Transformation diagrams above $T_g$.** The unit of entropy and volume are J/K/atom.g. and $m^3$ respectively.

| Crystal | lnA | $T_{0m}$ (K) | B (K) | $\varepsilon_{ls0}$ | $-\theta_{0g}$ | $-\theta_{0m}$ | $\varepsilon_{nm0}$ | $\Delta S_m$ | $V_m\, m^3$ | n | $V\, m^3$ | Ref |
|---|---|---|---|---|---|---|---|---|---|---|---|---|
| BMG N°1 | 86 | 447 | 4135 | 1.724 | 0.540 | 0.46 | 0.68 | 8.55 | $8\times10^{-6}$ | 13 | $20\times10^{-9}$ | [34,73] |
| BMG N°2 | 86 | 430 | 5170 | 1.718 | 0.545 | 0.464 | 0.69 | 8.74 | $8\times10^{-6}$ | 13 | $9.16\times10^{-9}$ | [35,74,75,83] |
| BMG N°4 | 84 | 472 | 5900 | 1.723 | 0.577 | 0.461 | 0.68 | 8.75 | $9.76\times10^{-6}$ | 13 | $3\times10^{-9}$ | [36,77-79] |

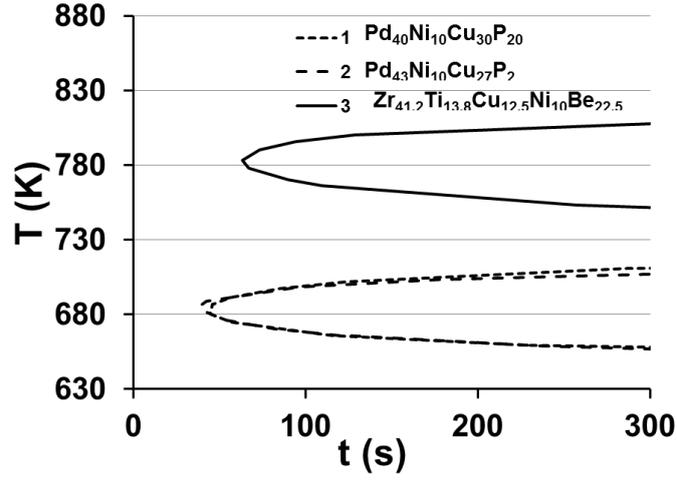

**Figure B1.** The calculated TTT diagrams of $Pd_{40}Ni_{10}Cu_{30}P_{20}$, $Pd_{43}Ni_{10}Cu_{27}P_{20}$, and $Zr_{41.2}Ti_{13.8}Cu_{12.5}Ni_{10}Be_{22.5}$. They are in agreement with the experimental observations of the crystallization nucleation.

A study of isothermal nucleation time is made after quenching the melt from above $T_m$ down to the annealing temperature. The time-lag $\tau^{ns}$ for transient nucleation has to be included in the calculation of the total nucleation time t which is equal to ($\tau^{ns} + \pi^2/6 \times t_{sn}$) [37]. The calculated liquid–crystal TTT diagrams of three glass-forming melts $Pd_{40}Ni_{10}Cu_{30}P_2$, $Pd_{43}Ni_{10}Cu_{27}P_{20}$, and $Zr_{41.2}Ti_{13.8}Cu_{12.5}Ni_{10}Be_{22.5}$ are represented in Figure B1 in agreement with the experimental studies [34-36]. The parameters given in Table B1 are used. The 13-atom supercluster is the minimum size entity which is formed by homogeneous nucleation. These superclusters are condensed when (B1) is respected with n = 13 inducing a spontaneous supercluster growth up the critical size and beyond it because the cluster energy barrier $\Delta G^*_{13}/k_BT$ is much larger than the critical value $\Delta G^*_{2ls}/k_BT$:

$$\ln(J_{13m} \times v \times t_{sn}) = \ln(K_{ls} \times v \times t_{sn}) - \frac{\Delta G^*_{13m}}{k_BT} + \frac{\Delta G_{13m}}{k_BT} = 0, \tag{B1}$$

where $\Delta G_{13m}$ and $\Delta G^*_{13m}/k_BT$ are given by (17) and (22) with n = 13. The 13-atom cluster nucleation rate $J_{13}$ is equal to $(v.t_{sn})^{-1}$ because $\ln(J_{13}.v.t_{sn}) = 0$, where v is the sample volume for crystallization and $t_{sn}$ the steady-state nucleation time of the 13-atom supercluster. The $\varepsilon_{13m0}$ and lnA values have been varied to reproduce the total nucleation time t and the nose temperatures of the experimental TTT diagrams [29,42].

**Acknowledgments**

The first steps of this work have been also sponsored by the sino-french Laboratory for the Application of Superconductors and Magnetic Materials involving the Northwestern Polytechnical University (NPU) in Xi'an, the Institut Polytechnique de Grenoble and the Centre National de la Recherche Scientifique in Paris. These initial works were presented at the Workshop of Solidification Processing-1: Melt Structure and Nucleation in 2011 in Xi'an (P. R. China), at the 5th International Workshop on Materials Analysis and Processing in 2012 in Autrans (France) and at the meeting of the Magneto-Science Society of Japan in 2012 in Kyoto. The author thanks Professor Lian Zhou and Professor Wanqi Jie in Xi'an, Professor Eric Beaugnon in Grenoble and Professor Tsunehisa Kimura in Kyoto for their invitations to present this controversial point of view on the vitreous transition thermodynamic origin. Thanks are due to Jean-Louis Soubeyroux for communication of many references.

**References**

[1] I. Cohen, A. Ha, X. Zhao, M. Lee, T. Fischer, M.J. Strouse, and D. Kivelson, A low temperature amorphous phase in a fragile glass-forming substance, J. Phys. Chem.100 (1996) 8518-8526.

[2] R. Kurita, H. Tanaka, Critical-like phenomena associated with a liquid-liquid transition in a molecular liquid, Science, 306 (2004) 845-848.

[3] H. Tanaka, R. Kurita, H. Mataki, Liquid-liquid transition in the molecular liquid triphenyl phosphite, Phys. Rev. Lett. 92 (2004) 02570.

[4] D. Wales, Energy landscapes. Applications to clusters, biomolecules and glasses, Cambridge U-K: Cambridge University Press, 2003.

[5] R. Doremus, Melt viscosities of silicate glasses, Am. Ceram. Soc. Bull, 82 (2003). 59-63..

[6] M. Ojovan, K. Travis and R. Hand, Thermodynamic parameters of bonds in glassy materials from viscosity-termperature relationships, J. Phys.: Condens. Matter, 19 (2007) 415107 (12 pp).

[7] M. I. Ojovan, Ordering and structural changes at the glass-liquid transition, J. Non-Cryst. Sol. 382 (2013) 79-86.

[8] N. N. Medvedev, A. Geiger and W. Brostow, Distinguishing liquids from amorphous solids: percolation analysis on the Voronoi network, J. Chem. Phys. 93 (1990) 8337-8342.

[9] Frank F. C., "Supercooling of liquids," Proc. R. Soc. Lond. A215 (1952) 43-46.

[10] A. V. Evteev, A. T. Kosilov and E. V. Levchenko, "Atomic Mechanisms of pure iron vitrification," J. Exp. Theor. Phys. 99 (2004) 522-529.

[11] Y.T. Shen,T.H. Kim, A.K. Gangopadhyay, K.F. Kelton, Icosahedral order, frustration, and the glass transition: evidence from time-dependent nucleation and supercooled liquid structure studies, Phys. Rev. Lett. 102 (2009) 057801.

[12] R. P. Wool, Twinkling fractal theory of the glass transition, J. Polym. Sci. B, 46 (2008) 2765-2778..

[13] R. P. Wool and A. Campanella, J. Polym. Phys. part B: 47 (2009) 2578-2589.

[14] L.Berthier, G. Biroli, J. Bouchaud, L. Cipelletti, D.E. Masri, D. L'Hôte, F. Ladieu, M. Pierno, Direct experimental evidence of a growing length scale accompanying the glass transition, Science, 310 (2005) 1797-1800.

[15] J. F. Stanzione III, K. E. Strawhecker and R. P. Wool, Observing the twinkling fractal nature of the glass transition, J. Non-Cryst. Sol. 357 (2011) 311-319.

[16] K.F. Kelton,G.W. Lee, A.K. Gangopaddhyay, R.W. Hyers, T.J. Rathz, J.R. Rogers, M.B. Robinson, and D.S. robinson, First-X-Ray scattering studies on electrostatically levitated metallic liquids: demonstrated influence of local icosahedral order on the nucleation barrier, Phys. Rev. Lett. 90 (2003) 195504.

[17] N.A. Mauro, J.C. Bendert,A.J. Vogt, J.M. Gewin, K.F. Kelton, High energy x-ray scattering studies of the local order in liquid Al, J. Chem. Phys. 135 (2011) p. 044502.

[18] A. Hirata, Y. Hirotsu, T. Ohkubo, T. Hanada, and V.Z. Bengus, Compositional dependence of local atomic structure in amorphous $Fe_{100-x}$ (x = 14, 17, 20) alloys studied by electron diffraction and high resolution electron microscopy, Phys. Rev. B, 74 (2006) 214206 (9 pp),

[19] T. Kirpatrick and D. Thirumalai, Mode-coupling theory of the glass transition, Phys. Rev. A, 31 (1985) 939-944.

[20] T. Kirpatrick and D. Thirumalai, Random first-order phase transition theory of the structural glass

transition, in: P. Wolynes and V. Lubchenko, (Eds.), Structural Glasses and supercooled liquids, , Wiley and sons, Hoboken, New Jersey, 2012, pp. 224-236.

[21] C. Angell, "Evidence for a new paradigm, and a new relation to strong liquids, in: P. Wolynes and V. Lubchenko,(Eds), Structural glasses and supercooled liquids, Wiley ans sons, Hoboken, New Jersey, 2012, pp. 237-278.

[22] D. S. Sanditov, "A criterion for the glass-liquid transition, J. Non-Cryst. Sol. 385 (2014) pp. 148-152.

[23] G.A. Almyras, Ch.E. Lekka, N. Mattern, G.A. Evangelakis, On the microstructure of the Cu65Zr35 and Cu35Zr65 metallic glasses, Scripta Materialia, 62 (2010) pp. 33-36.

[24] A. Berton, J. Chaussy, R. Rammal, J. Souletie, R. Tournier, Energy relaxation in the spin glass Au-Fe 4% at low temperatures, J. Physique- Lettres, 40 (1979) pp. L391-L394.

[25] W. Kauzmann, The nature of the glassy state and the behavior of liquids at low temperatures., Chem. Rev, 45 (1948) pp. 219-256.

[26] R.F.Tournier, Influence of Fermi energy equalization on crystal nucleation in glass melts, J. All. Comp. 483 (2009), pp. 94-96.

[27] D. T. Wu, L. Granazy and F. Spaepen, Nucleation and the solid-liquid interfacial free energy, MRS. Bull. December 2004, pp.945-950. [http://w.w.w.mrs.org/publications/bulletin].

[28] R.F. Tournier, Presence of intrinsic growth nuclei in overheated and undercooled liquid elements, Phys. B.: Condens. Matter, 392 (2007) pp. 79-91.

[29] R.F. Tournier, Crystal growth nucleation and Fermi energy equalization of intrinsic spherical nuclei in glass-forming melts, Sci. Technol. Adv. Mater, 10 (2009) p. 014607 (12pp).

[30] R. F. Tournier and E. Beaugnon, Texturing by cooling a metallic melt in a magnetic field, Sci. Technol. Adv. Mater, 10 (2009), p. 014501, (10pp).

[31] R.F. Tournier, Thermodynamic origin of the vitreous transition, Materials, 4 (2011) pp. 869-892.

[32] R.F. Tournier, Thermodynamic and kinetic origin of the vitreous transition, Intermetallics,30 (2012) pp. 104-110.

[33] R.F. Tournier, Thermodynamics of the vitreous transition, Revue de Metall. 109 (2012) pp. 27-33.

[34] J.F. Löffler, J. Schroers, W.L. Johnson, Time-temperature-transformation diagram and microstructures of bulk glass forming Pd40Cu30Ni10P20, Appl. Phys. Lett. 77 (2000) pp. 681-683.

[35] J. Schroers, Y. Wu, R. Busch, W.L. Busch, Transition from nucleation controlled to growth controlled crystallization in Pd43Cu27Ni10P20, Acta mater. 49 (2001) pp. 2773-2781.

[36] S. Mukherjee, Z. Zhou, J. Schroers, W.L. Johnson, and W.K. Rhim, Overheating threshold and its effect on time-temperature-transformation diagrams of zirconium based bulk metallic glasses, Appl. Phys. Lett. 84, (2004) pp. 5010-5012.

[37] I. Gutzow, J. Schmeltzer, The vitreous state, Ed: Springer-Verlag, Berlin Heidelberg New York, 1995.

[38] D. Turnbull, Kinetics of solidification of supercooled liquid mercury droplets, J. Chem. Phys. 20 (1952) pp. 411-424.

[39] D. Turnbull, J.C. Fisher, Rate of nucleation in condensed systems, J. Chem. Phys. 17 (1949) pp. 71-73.

[40] F. Faupel, W. Frank, M.-P. Macht, H. Mehrer, V. Naundorf, K. Rätzke, H. Schober, S. Sharma and H. Teichler, Diffusion in metallic glasses and supercooled melts, Rev. Modern Phys., 75 (2003) pp. 238-280.

[41] C.-Y. Liu, J. He, R. Keunings, C.C.-Y. Bailly, New linearized relation for the universal viscosity-termperature behavior of polymer melts, Macromolecules, 39 (2006) pp. 8867-8869.

[42] G. Adam, J.H. Gibbs, On the temperature dependence of cooperative relaxation properties in glass-forming liquids, J. chem. Phys. 43 (1965) pp. 139-146.

[43] H. Tanaka, Relationship among glass-forming ability, fragility and short-range bond ordering, J. Non-Cryst. Sol., 351 (2005) pp. 678-690.

[44] K.J. Dauwson, K.L. Kearns, L. Yu, W. Steffen, and M.D. Ediger, Physical vapor deposition as a route to

hidden amorphous states, PNAS, 106 (2009) pp. 15165-15170.

[45] I.K. Ishi, H. Nakayama, S. Hirabayashi, and R. Moriyama, Anomalously high-density glass of ethylbenzene prepared by vapor deposition at temperatures close to the glass-transition temperature, Chem. Phys. Lett., 459 (2008) pp. 109-112.

[46] V. Vinet, L. Magnusson, H. Frederiksson and P. J. Desre, Correlations between surface and interface energies with respect to crystal nucleation, J. Coll. Interf. Sci. 255 (2002) pp. 363-374.

[47] V.I. Kuzmin, D.L. Tytik, D.K. Belashchenko, A.N. Sirenko, Structure of silver clusters with magic numbers of atoms by data of molecular dynamics, Colloid J. 70 (2008) pp. 284-296.

[48] R. Tournier, Crystallization of supercooled liquid elements induced by superclusters containing magic atom numbers, Metals, 4 (2014) 359-387.

[49] J.H. Perezpezko, Nucleation in undercooled liquids, Mater. Sci. Eng. 65 (1984) pp. 125-135..

[50] C.A. Angell, Entropy and fragility in supercooling liquids, J. Res. Natl. Inst. Stand. Technol., 102 (1997) pp. 171-185.

[51] R.F. Tournier, Crystal growth nucleation and equalization of Fermi energies of intrinsic nuclei and glass-forming melts, J. Conf. Series,14 (2009) p. 012116.

[52]  Z. Cernosek, J. Holubova, E. Cernoskova, M. Liska, Enthalpic relaxation and the glass transition, J. Optoelectron. Adv. Mat. 4 (2002) pp. 489-503.

[53] B. Wunderlich, Study of the change in specific heat of monomeric and polymeric glasses during the glass transition, J. Phys. Chem. 64 (1960) pp. 1052-1056..

[54] S. Wei, L. Gallino, R. Busch, and C.A. Angell, Glass transition with decreasing correlation length during cooling of Fe50Co50 superlattice and strong liquids, Nature Phys. 7 (2011) pp. 178-182.

[55] P. Richet, Y. Bottinga, Thermochemical properties of silicate glasses and liquids, Rev. Geophys. 24 (1986) pp. 1-25.

[56] C.T. Moynihan, and S. Cantor, Viscosity and its temperature dependence in molten BeF2, J. Chem. Phys. 48 (1968) pp. 115-119.

[57] H.B. Ke, P. Wen,D.Q. Zhao, W.H. Zhang, Correlation between dynamic flow and thermodynamic glass transition in metllic glasses, Appl. Phys. Lett. 96 (2010) p. 251902.

[58] V.V. Brazhkin, A.G. Lyapin, S.V. Popova, Y. Inamura, H. Saitoh, W. Utsumi, Structural study of phase transformations in solid and liquid halogenides ZnCl2 and AlCl3 under high pressure, Pis'ma v Zh. Eksper. teoret. Fiz. 82 (2005) pp. 808-815.

[59] G. Ferlat, A.P. Seitsonen, M. Lazzeri, F. Mauri, Hidden polymorphs drive vitrification in B2O3, Nature Mater., 11 (2012) pp. 925-930.

[60] D.F. Weil, J.F. Stebbins, R. Hon, , I.S.E. Carmichael, The enthalpy of fusion of anorthite, Contr. Mineral. Petrol., 74 (1980) pp. 95-102.

[61] M.B.  Myers, E.J. Felty, Heats of fusion of As2S3, As2Se3, As2Te3 and Sb2S3, J. Electrochem. Soc. 117, (1970) pp. 818-820

[62] A. Tverjanovich, Calculation of viscositty of chalcogenide glasses near glass transition temperature from heat capacity or thermal expansion data, J. Non-Cryst. Soc. 298 (2002) pp. 226-231.

[63] J.F. Stebbins, I.S.E. Carmichael, I.K. Moret, Heat Capacities and entropies of silicate liquids and glasses, Contrib. Mineral. Petrol. 86, (1984) pp. 131-148.

[64] R. Busch, E. Bakke, and W.L. Johnson, Viscosity of the supercooled liquid and relaxation at the glass transition of the Zr46.75Ti8.25Cu7.5Ni10Be27.5 bulk metallic glass forming alloy, Acta Mater. 46 (1998) pp. 4725-4732.

[65] R. Busch, and W.L. Johnson, The kinetic glass transition of the Zr46.75Ti8.25Cu7.5Ni10Be27.5 bulk metallic glass former-supercooled liquids on a long time scale, Appl. Phys. Lett. 72 (1998) pp. 2695-2697.

[66] H.S. Chen, D. Trurnbull, Evidence of a glass-liquid transition in a gold-germanium-silicon alloy, J. Chem. Phys. 48 (1968) pp. 2560-2571.

[67] J.F. Stebbins, I.S.E. carmichael, D.F. Weill, The high temperature liquid and glass heat contents and the heats of fusion of diopside, albite, sanidine, and nepheline, Am. Mineral., 68 (1983) pp. 717-73.

[68] A. Sipp, Y. Bottinga, P. Richet, New high viscosity data for 3D network liquids and new correlations between old parameters, J. Non-Cryst. Sol. 288 (2001) pp. 166-174.

[69] D. Huang, G.B. McKenna, New insights into the fragility dilemna in liquids, J. Chem Phys. vol. 114, pp. 5621-5630, 2001.

[70] L.-M. Wang, C.A. Angell, R. Richert, Fragility and thermodynamics in nonpolymeric glass-forming liquids, J. Chem. Phys. 125 (2006) p. 074505.

[71] J. Wong, C.A. Angell, Glass structure by spectroscopy, New York: Ed: Dekker, 1976.

[72] I.-R. Lu, G. Wilde, G.P. Görler, Thermodynamic properties of Pd-based glass-forming alloys, J. Non-Cryst. Sol. 250-252, (1999) pp. 577-581.

[73] N. Nishiyama, M. Horino, O. Haruyama, and A. Inoue, Abrupt change in heat capacity of supercooled Pd-Cu-Ni-P melt during continuous cooling, Mater. Sci. Eng. A304-306 (2001) pp. 683-686.

[74] G.J. Fan, J.F. Löffler, R.K. Wunderlich, H.J. Fecht, Thermodynamics, enthalpy relaxation and fragility of the bulk glass-forming liquid Pd43Cu27Ni10P20, Acta Materialia, 52 (2004) pp. 667-674.

[75] G.J. Fan, H.J. Fecht, E.J. Lavernia, Viscous flow of the Pd43Cu27Ni10P20 bulk metallic glass-forming melt, Appl. phys. Lett. 84 (2004) pp. 487-489.

[76] Z. Evenson, R. Busch, Equilibrium viscosity, enthalpy recovery and free volume relaxation in a Zr44Ti11Ni10Cu10Be25 bulk metallic glass, Acta Materialia, 59 (2011) pp. 4404-4415.

[77] R. Busch, Y.J. Kim, W.L. Johnson, Thermodynamics and kinetics of the undercooled liquid and the glass transition of the Zr41.2Ti13.8Cu12.5Ni10Be22.5 alloy, J. Appl. Phys. 77 (1995) pp. 4039-4043.

[78] R. Busch, A. Masuhr, W.L. Johnson, Thermodynamics and kinetics of Zr-Ti-Cu-Ni-Be bulk metallic glass-forming liquids, Mater. Sci. Eng. A304-30 (2001) pp. 97-102.

[79] S. Mukherjee, J. Schroers, Z. Zhou, W.L. Jognson, W.-K. Rhim, Viscosity and specific volume of bulk glass-forming alloys and their correlation with glass-forming ability, Acta Mater. 52 (2004) pp. 3689-3695.

[80] G. Wilde, I.-R. Lu, R. Willnecker, Fragility, thermodynamic properties, and thermal stability of Pd-rich glass-forming liquids," Mater. Sci. Eng., A375-377 (2004) pp. 417-421.

[81] G. Wilde, G.P. Görler, R. Willnecker, G. Dietz, Thermodynamic properties of Pd40Ni40P20 in the glassy, luiquid, and crystalline states, Appl. Phys. Lett. 65 (1994) pp. 397-399.

[82] J.N. Mei, J.L. Soubeyroux, J.J. Blandin, J.S. Li, H.C. Kou, H.Z. Kou, , L. Zhou, Strutural relaxation of Ti40Zr25Ni8Cu9Be18 bulk metallic glass, J. Non-Cryst. Sol. 357 (2011) pp. 110-115.

[83] J. Schroers, On the formability of bulk metallic glass in its supercooled liquid state, Acta Mater. 56 (2008) pp. 471-478.

[84] A. Legg, J. Sxchroers, R. Busch, Thermodynamics, kinetics, and crystallizatiion of Pt57.3Cu14.7Ni5.3P22.8 bulk metallic glass, Acta Mater. 55 (2007) pp. 1109-1116.

[85] S.C. Glade, R. Busch, D.S. Lee, W.L. Johnson, R.K. Wunderlich, H.J. Fecht, Thermodynamics of Cu47Ti34Zr11Ni8, Zr 52.5Cu17.9Ni14.6Al10Ti5 and Zr57Cu15.4Ni12.6Al10Nb5, J. Appl. Phys. 87 (2000) pp. 7242-7248.

[86] Q.K. Jian, X.D. Wang, X.P. Nie, G.Q. Zhang, H. Ma, H.J. Fecht, J. Bendnarck, H. Franz, Y.G. Liu, Q.P. Cao. J.Z. Jiang, Zr-(Cu, Ag)-Al bulk metallic glasses, Acta Materialia, 56 (2008) pp. 1785-1796.

[87] I. Gallino, M.B. Shah, R. Busch, Enthalpy relaxation and its relation to the thermodynamics and crystallization of the Zr58.5Cu15.6Ni12.8Al10.3Nb2.8 bulk metallic glass-forming alloy, Acta materialia, 55 (2007) pp. 1367-1376.

[88] Q.G. Meng, S.G. Zhang, J.G. Li, X.F. Bian, Strong liquid behavior of Pr55Ni25Al20 bulk metallic glass, J. All. Comp. 431 (2007) pp. 191-196.

[89] J. Souletie, The glass transition: Dynamic and static scaling approach, J. Phys. France, 51 (1990) pp. 883-

898.

[90] H.S. Chen, A method for evaluating viscosities of metallic glasses from the rates of thermal transformations, J. Non-Cryst. Sol. 27 (1978) pp. 257-263.

[91] Q.K. Zhang, H. Hahn, Study of the kinetics of free volume in Zr 45Cu39.3Al7Ag8.7 bulk metallic glasses during isothermal relaxation by enthalpy relaxation experiments, J. Non-Cryst. Sol., 355 (2009) pp. 2616-2621.

[92] S.C. Glade, W.L. Johnson, The viscous flow of the Cu47Ti34Zr11Ni8 glass forming alloy, J. Appl. Phys. 87 (2000) pp. 7249-7251.

[93] R. Busch, W. Liu, W.L. Johnson, Thermodynamics and kinetics of the Mg65Cu25Y10 bulk metallic glass-forming liquid, J. Appl. Phys. 83 1998) pp. 4134-4141.

[94] F. Sommer, Thermodynamics of liquid alloys, Mater. Sci. Eng., A226-228 (1997) pp. 757-762, 1997.

[95] C. Fan, A. Inoue, Influence of liquid states on the crystallization process of nanocrystal-forming Zr-Cu-Pd-Al metallic glasses, Appl. Phys. Lett. 75 (1999) pp. 3644-3646.

[96] Z.P. Lu, Y. Li, and C.T. liu, Glass-forming tendency of bulk La-Al-Ni-Cu-(Co) metallic glass-forming liquids, J. Appl. Phys. 93 (2003) pp. 286-290.

[97] Z.P. Lu, X. Hu, Y. Li, Thermodynamics of La based La-Al-Cu-Ni-Co alloys studied by temperature modulated DSC, Intermetallics, 8 (2000) pp. 477-480.

[98] S.H. Zhou, J. Schmid, F. Sommer, Thermodynamic properties of liquid, undercooled liquid and amorphous Al-Cu-Zr and Al-Cu-Ni-Zr alloys, Thermochim. Acta, 339 (1999) pp. 1-9.

[99] Z. Cernosek, J. Holubova, E. Cernoskova, Kauzmann temperature and the glass transition, J. optoelectron. adv. mat. 7 (2005) pp. 2941-2944.

[100] A. Raemy, T.F. Schweizer, Thermal behaviour of carbohydrates studied by heat flow calorimetry, J. Therm. Anal. 28 (1983) pp. 95-108.

[101] Gangasharan and S.S.N. Murthy, Nature of the relaxation processes in the supercooled liquid and glassy states of some glassy carbohydrates, J. Phys Chem. 99 (1995) pp. 12349-12354.

[102] S.S. Chang, B.B. Bestul, Heat capacity and thermodynamic properties of o-terphenil crystal, glass, and liquid, J. Chem. Phys. 56 (1972) pp. 503-516.

[103] G.P. Johari, Heat capacity and entropy of an equilbrium liquid from Tg to 0 K, and examining the conjectures of an underlying thermodynamic transition, Chem. Phys. 265 (2001) pp. 217-231.

[104] C. Alba, L.E. Busse, D.J. List, and C.A. Angell, Thermodynamic aspects of the vitrification of toluene, and xylene isomers, and the fragility of liquid hydrocarbons, J. Chem. Phys. 92 (1990) pp. 617-624.

[105] C. Alba-Simionesco, J. Fan, and C.A. Angell, Thermodynamic aspects of the glass transition phenomenon, J. Chem. Phys. 110 (1999) pp. 5262-5272.

[106] B.C. Hancock, M. Parks, What is the true solubility advantage for amorphous pharmaceuticals, Pharm. Res., 17 (2000) pp. 397-403.

[107] Y. Aso, S. Yoshioka, S. Kojima, Relationship between the crystallization of amorphous nifedipine, phenobarbital, and flopropione, and their molecular mobility as measured by their enthalpy relaxation and (1)H NMR relaxations, J. Pharm. Sci., 89 (2000) pp. 408-416.

[108] N. Lebrun, J.C. van Miltenburg, Calorimetric study of maltitol: correlation between fragility and thermodynamic properties, J. All. Comp. 320 (2001) pp. 320-325.

[109] M. Hurtta, I. Pitkänen, Quantification of low levels of amorphous content in maltitol, Thermochim. Acta, vol. 419 (2004) pp 19-29..

[110] S.L. Shamblin, X. Tang, L. chang, B.C. Hancock, M.J. Pikal, Characterization of the time scales of molecular motion in pharmaceutically important glasses., J. Phys. Chem. B, 103 (1999) pp. 4113-4121.

[111] K.J. Crowley, G. Zografi, The use of thermal methods for predicting glass-former fragility," Thermochim. Acta, 380 (2001) pp. 79-93.

[112] H. Fujimori, H. Mizukami, and M.Oguni, Calorimetric study of 1,3-Diphenyl-1,1,3,3-tetramethyldisiloxane...," J.Non-Cryst. Sol. 204 (1996) pp 38-46.

[113] L.-M. Wang, V. Velikov, and C.A. Angell, Direct determination of kinetic fragility indices of glassforming liquids by differential scanning calorimetry: kinetic versu thermodynamic fragilities, J. Chem. Phys. 117 (2002) pp. 10184-10192.

[114] T. Hikima, . Okamoto, M. hanaya, M. Oguni, Calorimetric study of triphenylethene: observation of homogeneous-nucleation-based crystallization, J. Chem. Thermodyn. 30 (1998) pp. 509-523.

[115] I. Tsukushi, O. Yamamuro, T. Ohta, T. Matsuo, H. Nakano, , and Y. Shinota, A calorimetric study on the configurational enthalpy and low-energy excitation of ground amorphous solid and liquid-quenched glass of 1,3,5-tri-a-naphtylbenzene, J. Phys. Condens. Matter, 8 (1996) p. 245.

[116] K. Kishimoto, H. Suga, and S. Seki, Calorimetric study of the glass state. VIII ....of isopropylbenzene.," Bull. Chem. Soc. Jpn. 46 (1973) p. 3020-3031.

[117] H. Finke and J. Messerly, J. Chem. Therm., vol. 5, p. 5247, 1973.

[118] C.A. Angell, J.C. Tucker, Heat capacities and fusion entropies of the tetrahydrates of calcium nitrate, cadmium nitrate and magnesium acetate..., J. Phys. Chem. 78 (1974) pp. 278-281.

[119] Y. Roos, Melting and glass transitions of low molecular weight carbohydrates, Carbohydr. Res. 238 (1993) p. 39-48.

[120] S.S.N. Murthy, A. Pakairay, N. Arya, Molecular relaxation and excess of entropy in liquids and their connection to the structure of glass, J. Chem Phys.,10 (1995) pp. 8213-8220.

[121] J.E. Kunzler, W.F. Giauque, J. Am. Chem. Soc. 74 (1952) pp. 797-800.

[122] D. Beyer, The search for sulfuric acid octahydrate: experimental evidence. J. Chem. Phys. 107 (2003) p. 2025.

[123] S.S. Chang, J.A. Horman, B.A. Bestul, Heat capacities and related thermal thermal data for diethylphthalate crystal, glass, and liquid to 360 K. J. Res. Natl. Bur. Stand. 71 (1967) p. 293.

[124] S. Takahara, O. Yamamuro, and T. Matsuo, Calorimetric study of 3-bromopentane: Correlation between structural relaxation time and configurational entropy, J. Phys. Chem. 99 (1995) pp. 9589-9592.

[125] M. Mizukami, H. Fujimori, and M. Oguni, Glass transitions and the responsible molecular motions in 2-methyltetrahydrofuran, Prog. Theor. Phys. Suppl. 126 (1997) pp. 79-82.

[126] K. Takeda, O. Yamamuro, and H. Suga, Thermodynamic study of 1-butene., J. Phys. Chem. Sol., 52 (1991) pp. 607-615.

[127] J.G. Aston, H.L. Fink, A.B. Bestul, A.L. Pace, and G.J. Szasz, The heat capacity and entropy, heats of fusion and vaporization and vapor-pressure of 1-butene. J. Am. Chem. Soc. 68 (1946) pp. 52-62.

[128] O. Yamamuro, I. Tsukushi, A. Lindqvist, S. Takahara, M. Ishikawa, and T. Matsuo, J. Phys. Chem. B, 102 (1998) p. 1605-1609.

[129] D.R. Douslin, H.F. Huffman, Low temperature thermal data on the five isometric hexanes, J. Am. Chem. Soc. 68 (1946) p. 1704-1708.

[130] S. Takahara, O. Yamamuro, and H. Suga, Heat capacities and glass transition of 1-propanol and 3-methylpentane under pressure: new evidence for the entropy theory, J. Non-Cryst. Sol. 171 (1994) pp. 259-270.

[131] J. F. Counsell, E.B. Lees, J.F. Martin, Thermodynamic properties of organic ... 2-methylpropanol and pentanol, J. Am. Chem. Soc. A, 8 (1968) pp. 1819-1823.

[132] S.S. Chang, A.B. Bestul, Heat capacities of selenium crystal (trigonal), glass, and liquid from 5 to 360 K.," J. Chem. Thermodyn. 6 (1974) pp. 325-344.

[133] H. Hikawa, M. Oguni, and H. Suga, Construction of an adiabatic calorimeter for a vapor deposited sample and thermal characterization of amorphous butyronitrile, J. Non-Cryst. Sol. 101 (1988) pp. 90-100.

[134] K. Naito, A. Mura, Molecular design for nonpolymeric organic dye glasses with thermal stability; relations between thermodynamic parameters and amorphous properties, J. Phys. Chem. 97 (1993) pp.

6240-6248.

[135] O. Haida, H. Suga, and S. Seki, Calorimetric study of the glassy state XII. Plural glass-transition phenomena of ethanol, J. Chem. Thermodyn. 9 (1977) pp. 1133-1148.

[136] G. Carlson, E.F. Westrum, Methanol: heat capacity, enthalpies of transition and melting and thermodynamic properties from 5-300 K, J. Chem. Phys. 54 (1971) p. 1464.

[137] C.A. Angell, D.L. Smith, Test of the entropy basis of the Vogel-Tammann-Fulcher equation- dielectric relaxation of polyalcohols near Tg, J. Phys. Chem. 86 (1982) pp. 3845-3852.

[138] K. Takeda, O. Yamamuro, T. Tsukushi, H.M. Suga, Calorimetric study of ethylene glycol and 1,3 propanediol: configurational entropy in supercooled polyalcohols, J. Mol. Struct. 479 (1999) pp. 227-235.

[139] C.A. Angell, E. Williams, K.J. Rao, and J.C. Tucker, Heat capacity and glass transition thermodynamics for zinc chloride, a failure of the first Davies-Joenes relation for dTc/dp, J. Phys Chem. 81 (1977) pp. 238-243.

[140] N.E. Shmidt, "Heat capacity and heat of fusion of crystalline boron oxide, Russ. J. Inorg.Chem.Engl. Trans. 11 (1966) pp. 241-247.

[141] C.T. Moynihan, S.N. Crichton, and S.M. Opalka, Linear and non-linear structural relaxation, J. Non-Cryst. Sol. 131-133 (1991) pp. 420-434.

[142] G.J. Fan, H. Choo, P.K. Liaw, Fragility of metallic glass-forming liquids: a simple thermodynamic connection, J. Non-Cryst. Sol. 351 (2005) pp. 3879-3883.

[143] M. Oguni, H. Hikawa, and H. Suga, Enthalpy relaxationin vapor-deposited butyronitrile, Thermochim. Acta. 158 (1990) pp. 143-156.

[144] R. F. Tournier, "Nucleation of crystallization in titanium and vitreous state in glass-forming melt," in H. Chang, Y. Lu, D. Xu, L. Zhou, (Eds), Ti-2011, Proceedings of the 12th WorldConference on Titanium, Beijing, China, (2012), 2, (2012) 1527-1531. ISBN: 978-7-03-033895-2.

[145] X.F. Bian, B.A. Sun, L.N. Hu, and Y.B. Jia, Fragility of superheated melts and glass-forming ability in Al-based alloys, Phys. Lett. A., 335 (2005) pp. 61-67.

[146 A. Inoue, W. Zhang, T. Zhang, and K. Kurokasa, High-strength Cu-Based bulk glassy alloys in Cu-Zr-Ti and Cu-Hf-Ti ternary alloys., Acta Mater. 49 (2001) pp. 2645-2652.

[147] N. Lebrun, J.C. van Miltenburg, O. Bustin, and M. Descamps, Calorimetric fragility of maltitol, Phase trans. 76 (1972) pp. 841-846.

[148] K.L. Kearns, K.R. Whitaker, M.D. Ediger, H. Huth, and C. Schick, Observation of low heat capacities for vapor-deposited glasses of indomethacin as determined by AC nanocalorimetry, J. Chem. Phys. 133 (2010) p. 014702 (10 p).

[149] K.L. Kearns, S.F. Swallen, M.G. Ediger, T. Wu, Y. Sun, and L. Yu, Hiking down the energy landscape: Progress toward the Kauzmann temperatture via vapor deposition, J. Phys. Chem. B, 112 (2008) pp 4934-4942.

[150] K.L. Kearns, S.F. Swallen, M.G. Ediger, Y. Sun, Calorimetric evidence for two distinct molecular packing arrangements in stable glass of indomethacin, J. Phys. Chem. B, 113 (2009) pp 1579-1586.